\documentclass[aps,prd,twocolumn,
               notitlepage,mathrsfs,
               amssymb,amsmath,amsfonts,
               superscriptaddress,
               longbibliography,
               nofootinbib,floatfix]{revtex4-2}

\usepackage{graphicx}
\usepackage{url}
\usepackage[linktocpage,breaklinks]{hyperref}
\usepackage[capitalize]{cleveref}
\usepackage[usenames,dvipsnames]{xcolor}
\hypersetup{colorlinks=true,
            citecolor=NavyBlue,
            linkcolor=magenta,
            urlcolor=magenta}
\usepackage{multirow,array}
\DeclareMathAlphabet{\pazocal}{OMS}{zplm}{m}{n}
\usepackage{journals}

\begin{document}

\title{Black hole-accretion disk collision in general relativity: Axisymmetric simulations}

\date{\today}

\author{Alan Tsz-Lok Lam}
\email{tszlok.lam@aei.mpg.de}
\affiliation{Max-Planck-Institut f\"ur Gravitationsphysik (Albert-Einstein-Institut), Am M\"uhlenberg 1, D-14476 Potsdam-Golm, Germany}

\author{Masaru Shibata}
\affiliation{Max-Planck-Institut f\"ur Gravitationsphysik (Albert-Einstein-Institut), Am M\"uhlenberg 1, D-14476 Potsdam-Golm, Germany}
\affiliation{Center of Gravitational Physics and Quantum Information, Yukawa Institute for Theoretical Physics, Kyoto University, Kyoto, 606-8502, Japan} 

\author{Kyohei Kawaguchi}
\affiliation{Max-Planck-Institut f\"ur Gravitationsphysik (Albert-Einstein-Institut), Am M\"uhlenberg 1, D-14476 Potsdam-Golm, Germany}
\affiliation{Center of Gravitational Physics and Quantum Information, Yukawa Institute for Theoretical Physics, Kyoto University, Kyoto, 606-8502, Japan} 

\author{Joaquin Pelle}
\affiliation{Max-Planck-Institut f\"ur Gravitationsphysik (Albert-Einstein-Institut), Am M\"uhlenberg 1, D-14476 Potsdam-Golm, Germany}

\begin{abstract}
Motivated by recent discoveries of X-ray quasi-periodic eruptions, we revisit the collision of a black hole and an accretion disk.
Assuming that they are orbiting a supermassive black hole in orthogonal orbits, we perform a general relativistic simulation of the collision, varying the relative velocity $V_0$ from $0.032c$ to $0.2c$ (where $c$ is the speed of light) with a variety of disk thickness and a realistic local density profile for the disk.
Our findings indicate that the mass of the outflow matter from the disk, $m_{\rm ej}$, is slightly less than the expected value.
Meanwhile, the typical energy associated with this outflow $E_{\rm ej}$ is $\sim m_{\rm ej}V_0^2$.
Thus, the predicted peak luminosity from disk flares is approximately equal to the Eddington luminosity of the black hole, whereas the peak time and duration of the flares, which are $\propto m_{\rm ej}^{1/2}$, are shorter than that previously believed.
We also demonstrate that the property of the outflow matter induced by the incoming and outgoing stages of the black hole collision is appreciably different.
We find that a high mass accretion rate onto the black hole from the disk persists for a timescale of $\sim 10^6$ Schwarzschild time of the black hole after the collision for $V_0/c \lesssim 0.1$, making this long-term accretion onto the black hole the dominant emission process for black hole-disk collision events.
Implications of these results are discussed.  
\end{abstract}

\maketitle

\section{Introduction}

The collision between a star and a disk orbiting a supermassive black hole at an extragalactic center can induce a bright electromagnetic transient~\cite{1998ApJ...507..131I, 2023ApJ...957...34L, 2025ApJ...978...91Y, 
dai2010quasi, xian2021x, tagawa2023flares, zhou2024probing, linial2024ultraviolet, vurm2024radiation}.
Recent discoveries of X-ray quasi-periodic eruption (QPE) \cite{2019Natur.573..381M, 2020A&A...636L...2G, 2021Natur.592..704A, 2024ApJ...965...12C}, some of which are likely to be induced by the collision between a star and a disk formed after tidal disruption of another star by a supermassive black hole (or a newly-formed dense disk around a supermassive black hole by an unknown mechanism)~\cite{2024Natur.634..804N,  2023A&A...675A.152Q, Chakraborty:2025ntn, Hernandez-Garcia:2025ruv}, illustrate that this type of collision can be a powerful source of luminous transients. 

Because it is natural to believe the presence of numerous black holes orbiting supermassive black holes near the galactic centers, we can speculate that collisions between a black hole and an accretion disk occur at a certain rate.
Such collisions have already been investigated through numerical simulations as noted in \cite{1998ApJ...507..131I}, which suggest that these events can produce bright transients~\cite{lehto1996oj, pihajoki2016black, franchini2023quasi}.
The peak luminosity for this type of transients is expected to reach the Eddington luminosity of the black hole given by $L_{\rm Edd} \approx 1.4 \times 10^{39}(M_{\rm BH}/10M_\odot)$\,erg/s with $M_{\rm BH}$ being the black hole mass.
Therefore, the luminosity is generally not sufficient to allow for the observation of extragalactic events unless the black hole mass is significantly large.
Nevertheless, this phenomenon can be utilized for exploring intermediate-mass black holes with a mass $\gg 10M_\odot$ (see, e.g., \cite{2024arXiv241012090Z}).
Additionally, the collision between a supermassive black hole and an accrete disk may serve as a luminous optical transient source in binary supermassive black hole systems~\cite{2008Natur.452..851V} (but see \cite{2023MNRAS.522L..84K}).
The late inspiral phases of an intermediate-mass or supermassive black hole orbiting another supermassive black hole with mass $10^5$--$ 10^7 M_\odot$ are among the most important targets for the Laser Interferometer Space Antenna (LISA) \cite{2017arXiv170200786A}.
In the hypothetical presence of this LISA source, it is reasonable to consider that the secondary could shine due to collisions with a disk formed transiently before the system evolves into a compact orbit suitable for LISA detection.
These considerations motivate us to conduct a detailed theoretical study of the black hole-disk collisions.

Recent numerical simulations have examined the interaction between an orbiting star or a small black hole and the geometrically thick accretion disk surrounding a supermassive black hole ~\cite{sukova2021stellar, pasham2024case, ressler2024black}.
These studies consistently show that a Bondi-Hoyle-Lyttleton-type flow (e.g., \cite{hoyle1939effect, bondi1944mechanism, 1989ApJ...336..313P, 1998ApJ...494..297F, edgar2004review}) develops around the orbiting object, and that mildly quasi-periodic outflows can be launched from the system.
Furthermore, in the case of a secondary black hole with a mass of $10\%$ of the primary black hole, spin-orbit coupling can induce long-term precession of the disk and jet \cite{ressler2024black}.

This paper pays particular attention to the collision between a black hole and a geometrically thin disk.
Although previous work~\cite{1998ApJ...507..131I} provided valuable insights into this topic, there are several areas that warrant further exploration, listed as follows;
(i) disks with uniform density were considered in a previous study~\cite{1998ApJ...507..131I}, but in reality, a disk orbiting a central supermassive black hole should have a nonuniform density profile along the direction perpendicular to the orbiting plane;
(ii) non-relativistic simulations were performed despite the fact that the black hole can travel at a high velocity of the order of 10\% of the speed of light $c$;
(iii) the mass and energy of the outflow driven by the collision, which are crucial for predicting electromagnetic signals, were not thoroughly analyzed;
(iv) the study only focused on a specific black hole velocity of $\approx 3c/80$, possibly due to the computational limitation at that time, although the collision process may depend quantitatively on the relative velocity; (v) the duration of the simulations was not sufficiently long to explore the long-term accretion process onto the black hole following the collision.

Therefore, in this paper, we systematically investigate the physical consequence from the collision between a black hole and a disk using a new fixed mesh-refinement (FMR) hydrodynamics code {\tt SACRA-2D}~\cite{Lam:2025pmz}.
Each simulation is performed for a sufficiently long time to clarify the long-term mass accretion process onto the black hole.
The FMR algorithm is well-suited for this problem because the thickness of the disk is significantly larger than the Schwarzschild radius of the colliding black hole, denoted as $r_{\rm H}$, and we have to follow the evolution of the system over a wide dynamical range; i.e., the distance between the black hole and the disk varies from $\gg r_{\rm g}(=GM_{\rm BH}/c^2)$ to $\sim r_{\rm g}$ and again to $\gg r_{\rm g}$ during the collision process.
We perform the simulations by setting up a plausible disk model for various relative velocities between the black hole and the disk until the matter accretion onto the black hole becomes inactive.
In addition, we carefully analyze the properties of the outflow matter arising from the collision, as well as the mass swallowed into the black hole.
Ultimately, our findings suggest that the long-term mass accretion onto the black hole after the collision is likely the primary source of bright emissions, rather than the collision itself.

The paper is organized as follows. \cref{sec2} describes the setup of the simulation.
After summarizing the diagnostics in \cref{sec3}, in \cref{sec4}, we present the numerical results, paying particular attention to the properties of the matter outflowed from the disk in the collision and to the long-term mass accretion rate onto the black hole.
In \cref{sec5}, implications of the present results are discussed, and then, we summarize the paper in \cref{sec6}.   
Throughout this paper, $c$ and $G$ denote the speed of light and gravitational constant, respectively, while the Greek letter $\mu$ denotes the spacetime coordinates.
We denote the gravitational radius of the black hole by $r_{\rm g}:=GM_{\rm BH}/c^2$, and $t_{\rm g}:=r_{\rm g}/c$. 

\section{Setup}\label{sec2}

We consider the collision between a black hole of $M_{\rm BH}$ orbiting a more massive (primary) black hole of mass $M_{\rm c} \gg M_{\rm BH}$ and an accretion disk around the primary black hole in a {\em local} approximation.
The profile of the disk is determined by the gravity of the primary black hole, with its self-gravity neglected.
The orbital radius of the (secondary) black hole around the primary black hole is denoted as $R$, which is assumed to be much larger than the gravitational radius of the primary black hole $R_{\rm g}:=GM_{\rm c}/c^2$, while the thickness of the disk $H$ is assumed to be smaller than $R$.
In the present {\em local} approximation, we set the secondary black hole moving along the $z$-direction and a disk of uniform density on the disk plane (i.e., in the perpendicular direction to the secondary black hole motion).
The asymptotic relative velocity is set as $V_0$, which is assumed to be $\sqrt{GM_{\rm c}/R}$; for simplicity, the secondary black hole is assumed to be in a circular orbit around the primary black hole with the orbit perpendicular to the disk plane.
Then, the orbital period of the secondary black hole around the primary is written as
\begin{align}
\begin{split}
P_{\rm orb} &= 2\pi\sqrt{\frac{R^3}{GM_{\rm c}}}
=2\pi \left( \frac{V_0}{c} \right)^{-3} \left(\frac{M_{\rm c}}{M_{\rm BH}}\right) t_{\rm g} \\
&\approx 5\times 10^6 
\left(\frac{V_0}{0.05c}\right)^{-3}
\left(\frac{Q}{100}\right) t_{\rm g} \,,
\end{split}
\end{align}
where $Q:=M_{\rm c}/M_{\rm BH}$ is the mass ratio.
Our {\em local} simulation is valid only in $\sim P_{\rm orb}/4$, i.e., in a timescale $\sim 10^6t_{\rm g}$ assuming a mass ratio of $Q \sim 100$.
This model may not be very reliable for longer timescale evolution with $t \agt P_{\rm orb}/4$.

In the actual numerical computation, we fix the secondary black hole at the origin and consider a disk initially located far from the black hole, with a uniform velocity in the $-z$ direction toward the black hole.
The velocity in other directions is not considered for the disk for simplicity.
General relativistic simulations are performed in a fixed background of the black hole, neglecting the self-gravity of the disk, but the gravity from the primary black hole is phenomenologically taken into account in the hydrodynamics equation (see \cref{eq6}).
The Kerr-Schild coordinates are used following \cite{1998ApJ...507L..67F}, but the spin parameter of the black hole is set to zero in this work.
For the simulations, we employ our axisymmetric FMR code {\tt SACRA-2D} (see \cite{Lam:2025pmz} for the details of this code). 

When the distance between the secondary black hole and the disk is large enough, the disk can be considered to be locally in force balance between the pressure and the gravity from the primary black hole.
The force balance along the cylindrical radial direction determines the Kepler motion with orbital angular velocity $\Omega=\sqrt{GM_{\rm c}/R^3}$, and consequently the relative velocity $V_0=R\Omega$.
On the other hand, the force balance for the direction perpendicular to the disk can be approximately written as
\begin{align} \label{eq1}
\frac{1}{\rho}\frac{dP}{dz}=-\frac{GM_{\rm c}}{R^3}(z-z_0)=-\Omega^2 (z-z_0), 
\end{align}
where $\rho$ and $P$ are the rest-mass density and the pressure, respectively, and $z_0$ denotes the momentary location of the maximum density in the disk.
We assume Newtonian gravity for simplicity in determining the disk structure.
It should be noted that in a realistic situation, the disk has the orbital velocity as well as the density and velocity gradients along the disk plane, but we ignore such effects in the present study to impose axial symmetry. 

In this paper, we also assume that the pressure is dominated by the radiation pressure and initially use the equation of state of the form
\begin{align} \label{eq2}
P=K\rho^{4/3}, 
\end{align}
where $K$ is the polytropic constant.
The corresponding speed of sound is approximately written as $c_{\rm s} \approx \sqrt{4P/3\rho}$ if $c_{\rm s}$ is much smaller than the speed of light, and thus, in the following, we choose $K$ with which the sound speed of the disk $c_{\rm s}$ is much smaller than the relative velocity $V_0$.
During the numerical evolution of the system, we use the $\Gamma$-law equation of state with $\Gamma=4/3$, i.e., $P=\rho \varepsilon/3$, with $\varepsilon$ being the specific internal energy.
We implicitly assume that most of the region inside the disk is optically thick and radiative cooling is inefficient within the dynamical timescale of the black hole-disk collision in this problem, while the diffusion timescale is short enough to form a geometrically thin disk, which is indeed the case for the assumed column density of the disk (see discussion in \cref{eq:sigma}).

With the given equation of state in \cref{eq2}, \cref{eq1} is integrated in the rest frame of the secondary black hole to give
\begin{align} \label{eq4}
\frac{4P}{\rho}=\frac{\Omega^2}{2}\left[ H^2- w_0^2 (z-z_0)^2 \right],
\end{align}
where $w_0:=[1-(V_0/c)^2]^{-1/2}$ is the Lorentz factor of the disk, $H$ denotes the thickness of the disk, and thus, the rest-mass density $\rho$ and pressure $P$ vanish for $|z-z_0|> H/w_0$. 
Then, the density profile of the disk for $|z-z_0|<H/w_0$ is derived as
\begin{align}
\rho &= \rho_0 \left[ 1-\frac{w_0^2 (z-z_0)^2}{H^2}\right]^3,
\end{align}
where $\rho_0$ denotes the maximum density, for which we consider $10^{-8}$--$10^{-9}\,{\rm g/cm^3}$ supposing that the disk is formed from a solar-type star tidally disrupted by a supermassive black hole of $M_{\rm c}\sim 10^5$--$10^7M_\odot$.
Since we assume that the self-gravity is negligible for the disk, $\rho_0$ could be arbitrarily chosen.
The column density $\Sigma$ is defined by
\begin{align} \label{eq:sigma}
\Sigma &= \int_{-H/w_0+z_0}^{H/w_0+z_0} \rho w_0 \, dz=\frac{32}{35}\rho_0 H.
\end{align}
We broadly assume that $\Sigma$ is of order of $M_\odot/(10R)^2 \sim 10^4(M_{\rm c}/10^6M_\odot)^{-2}(100R_{\rm g}/R)^2\,{\rm g/cm^2}$.
Therefore, the disk is assumed to be optically thick to the Thomson scattering for which the opacity $\kappa$ is $\approx 0.35\,{\rm cm^2/g}$ for the fully ionized gas composed of hydrogen and helium. 

In the simulations, since the density structure of the disk has to be preserved at least approximately before colliding with the black hole, we phenomenologically add a force term to the Euler equation.
Specifically, the force term associated with the lapse function, $\alpha$, is modified to be
\begin{align} \label{eq6}
- \psi^{6} \rho h w^2 \partial_z \alpha \rightarrow - \psi^{6} \rho h w^2 \left[ \partial_z \alpha + c^{-2}\alpha \Omega^2 \bar z f(\bar z) \right],
\end{align}
in the source term $s_{S_z}$ in Eq.~(21d) of \cite{Lam:2025pmz},
where $\bar z:= z - z_0 + V_0 t$ is the vertical distance from the disk center, $\alpha$ and $\psi$ are, respectively, the background lapse function and conformal factor of the secondary black hole with mass $M_{\rm BH}$, $h$ is the specific enthalpy, $w:=\alpha c u^t$ is the Lorentz factor of the fluid element with $u^\mu$ being the corresponding four-velocity, and $f(z)$ is a correction factor
\begin{align}
f(\bar z)={\rm min}(1, z_{\rm cut}^3/\bar{z}^3),    
\end{align}
where the gravity cutoff $z_{\rm cut}$ is a constant parameter for smearing out the additional gravity effect far away from the disk with $z_{\rm cut}=10H$ as fiducial but varying in a range $5H$--$40H$.
Note that a factor of background lapse function $\alpha$ in the second term of the right-hand side of \cref{eq6} is added to the additional force term to suppress its effect under the influence of strong gravity near the secondary black hole, while the correction still reduces to the form of Newtonian gravity asymptotically. 

Using the relation $V_0^2=GM_{\rm c}/R$, the ratio $z_{\rm cut}/R$ can be expressed as 
\begin{equation}
\frac{z_{\rm cut}}{R} =\frac{1}{4}
\left(\frac{z_{\rm cut}}{10^4\,r_{\rm g}}\right)\left(\frac{V_0}{0.05c}\right)^2\left(\frac{Q}{100}\right)^{-1}. 
\end{equation}
We suppose the parameters satisfy $z_{\rm cut}/R\alt 1$.
Typically, we consider a mass ratio $Q \sim 100$, and for lower values of $z_{\rm cut}/r_{\rm g}$ we expect lower values of $Q$.

The accretion radius of the black hole $r_a$ is defined by
\begin{align}
r_{\rm a}:=\frac{2GM_{\rm BH}}{V_0^2}. 
\end{align}
As the units of the accreted mass and mass accretion rate, we often refer, respectively, to
\begin{align}
m_{\rm a}&:= \pi r_{\rm a}^2 \Sigma \approx 4.4 \times 10^{28} M_{\rm BH,4}^2 v_{0.05}^{-4} \Sigma_4\,\,{\rm g}, \\
\dot m_{\rm a}&:= \pi r_{\rm a} V_0 \Sigma \approx 5.6\times 10^{25}M_{\rm BH,4} v_{0.05}^{-1} \Sigma_4\,\,{\rm g/s},\label{dotma}
\end{align}
where $M_{\rm BH,4}:=M_{\rm BH}/(10^4M_\odot)$, $v_{0.05}:=V_0/(0.05c)$, and $\Sigma_4:=\Sigma/(10^4\,{\rm g/cm^2})$.
Note that $\dot m_{\rm a}$ does not imply $dm_{\rm a}/dt$.
In this paper, we often choose $M_{\rm BH} \sim 10^4M_\odot$ as typical mass because the Eddington luminosity of such an intermediate-mass black hole is comparable to the peak luminosity of the observed QPEs~\cite{2019Natur.573..381M, 2020A&A...636L...2G, 2021Natur.592..704A, 2024ApJ...965...12C}.
The Eddington accretion rate is defined by
\begin{align} \label{eq:Eddington_acc}
\dot m_{\rm Edd}:=10L_{\rm Edd}c^{-2} \approx 1.5 \times 10^{22}M_{\rm BH,4}\,{\rm g/s}. 
\end{align}
Therefore, for a high value of the column density or slow black hole velocity, $\dot m_{\rm a}$ far exceeds the Eddington mass accretion rate as
\begin{equation}
\dot m_{\rm a} \approx 3.7 \times 10^3 v_{0.05}^{-1}\Sigma_4 \,\dot m_{\rm Edd}. 
\end{equation}

In this problem, the free parameters are the velocity $V_0$, the thickness of the disk represented as $H/r_{\rm a}$ (or equivalently $H/r_{\rm g}$), and the gravity cutoff expressed as $z_{\rm cut}/H$.
To explore the effects of these parameters, we conduct a series of simulations, systematically varying each one.
We initially position the center of the disk at $z_0=2000\,r_{\rm g}$, which is always larger than $r_{\rm a}$, and assign the velocity uniformly as $-V_0$ in the $z$ direction ($v^z=u^z/u^t=-V_0$).
The sound speed of the disk is initially set as $c_{\rm s}=c_{\rm s,max}=10^{-3}c$ at the disk center.
Assuming $P\approx aT^4/3$ where $a$ and $T$ are the radiation constant and temperature, the maximum temperature is estimated by 
\begin{align}
T \approx 2.3 \times 10^5 \left(\frac{\rho}{10^{-8}\,{\rm g/cm^3}}\right)^{1/4}
\left(\frac{c_{\rm s}}{10^{-3}c}\right)^{1/2}\,{\rm K}.
\end{align}
From \cref{eq4}, we can also express the ratio $H/R$ as
\begin{equation}\label{eq16}
\frac{H}{R}\approx 0.1 \,v_{0.05}^{-1}
\left(\frac{c_{\rm s,max}}{10^{-3}c}\right).
\end{equation}
Numerical results depend only weakly on the choice of maximum sound speed $c_{\rm s,max}$ as far as the condition of $V_0 \gg c_{\rm s,max}$ is satisfied.
However, if $c_{\rm s,max}$ approaches $V_0$, the post-collision evolution process is modified significantly.

We employ the two-to-one FMR structure in the computational domain, which is composed of a hierarchy of nested concentric grids overlaying on top of each other~\cite{Lam:2025pmz}.
The computational domain covers the region of $[0:x_{\rm max}]$ and $[-z_{\rm max}:z_{\rm max}]$ for $x$ and $z$, respectively.
It consists of $L$ levels of FMR domains, each of which contains an even number of grids $N$ and $2N$ in the $x$ and $z$ directions, with the grid spacing written as
\begin{align}
\begin{split}
    \Delta x^{(0)} &= x_{\max} / N, \\
    \Delta x^{(l)} &= \Delta x^{(l-1)} / 2, 
\end{split}&
\begin{split}
        \Delta z^{(0)} &= z_{\max} / N, \\
    \Delta z^{(l)} &= \Delta z^{(l-1)} / 2, 
\end{split}
\end{align}
for $l=1,2,\cdots,L-1$, where we choose $x_{\max}=z_{\max}$. Levels $0$ and $(L-1)$ represent the coarsest and finest levels, respectively.
The hydrodynamics variables are assigned at cell-centered positions.
We choose $L=14$, $N=128$, and $x_{\rm max}=z_{\rm max}=2.56\times 10^4\,r_{\rm g}$ for the present work, and adopt the HLLC Riemann solver \cite{Mignone:2005ft,White:2015omx,Kiuchi:2022ubj} for the evolution of the hydrodynamics variables.
In this setting, the grid spacing in the finest domain is $0.0244 \,r_{\rm g}$. We note $r_{\rm H}=2r_{\rm g}$ in the present coordinate choice of the black hole spacetime.  
We confirmed that with these settings, the numerical results are well-converged (see \cref{app1} for the convergence test with $N=64$ and $96$ and \cref{app2} for the comparison of Riemann solvers).
Numerical simulations were conducted on the Sakura cluster at the Max Planck Computing and Data Facility, utilizing an MPI configuration of $[4 \times 8]$ along with 5 OpenMP threads (see \cite{Lam:2025pmz} for further details).
Under this configuration, a simulation lasting for $t = 10^6 t_{\rm g}$ required approximately 20k CPU hours.

\section{diagnostics}\label{sec3}

We analyze the following quantities during each simulation: the mass accretion rate onto the black hole $\dot m_{\rm BH}$; the mass of the matter outflowed from the disk $m_{\rm ej}$; the outflow mass spectrum with respect to the velocity $dm_{\rm ej}/dv$ where the velocity $v$ is defined from the Lorentz factor in the comoving frame of the disk, $\tilde w$, by $v=c\sqrt{1-\tilde w^{-2}}$ (see \cref{eq:lorentz_factor} for the definition of $\tilde w$).
The total rest mass that falls into the black hole is calculated by
\begin{align}
\Delta m_{\rm BH}(t)=\int_0^t dt' \dot m_{\rm BH}(t').
\end{align}
The mass accretion rate onto the black hole is calculated by the surface integral of $-\rho u^r \sqrt{-g}$ where $g$ denotes the determinant of the spacetime metric. 

From $m_{\rm ej}$ and $dm_{\rm ej}/dv$ we evaluate the outflow energy $E_{\rm ej}$ by 
\begin{align}
E_{\rm ej}=\int \frac{c^2}{\sqrt{1-v^2/c^2}} \frac{dm_{\rm ej}}{dv} dv-m_{\rm ej}c^2. 
\end{align}
Since the majority of the ejecta has a velocity $v/c < 0.5$ in the present study, we can approximate it as
$E_{\rm ej}$ by 
\begin{align}
E_{\rm ej}=\frac{1}{2}\int v^2 \frac{dm_{\rm ej}}{dv} dv. 
\end{align}
This quantity is evaluated in the late time of the outflow evolution, for which the kinetic energy is much larger than the internal energy. The average velocity is also defined by $v_{\rm ave}:=\sqrt{2E_{\rm ej}/m_{\rm ej}}$.

The outflow mass is defined by the fluid element that satisfies the Bernoulli criterion $h \bar{u}_t < - c^3$,
where
\begin{align}
    \bar{u}_\mu = \Lambda_\mu{}^\nu u_\nu
\end{align}
is the four-velocity with respect to the rest frame of the disk, and $\Lambda_\mu{}^\nu$ is the Lorentz transformation matrix.
We obtain $\tilde w$ from
\begin{equation} \label{eq:lorentz_factor}
    \tilde w = - \frac{h \bar u_t}{c^3}.
\end{equation}
In addition to the Bernoulli criterion, we further impose a condition for the outflow by requiring the velocity of the outflow matter in the comoving frame of the disk, $v$, to be larger than $V_0$.
This is because the black hole and disk are, in reality, orbiting a supermassive black hole.
Thus, if the velocity of the matter is smaller than $V_0$, such components would return to the disk or fall toward the primary massive black hole.
It should also be mentioned that the majority of the second outflow components with the velocity of $\alt V_0$ are likely to be trapped by the secondary black hole.

\section{Result}\label{sec4}

\begin{table}
\centering
\caption{Parameters for models studied in this paper.
The columns of the table, from left to right, show the model names, the values of velocity $V_0/c$, disk thickness $H$ in units of $r_{\rm g}$ and $r_{\rm a}$, and gravity cutoff $z_{\rm cut}/H$.
The model names reflect $V_0/c$, $H/r_{\rm g}$, and $z_{\rm cut}/H$.
The initial location of the disk center is $z_0=2000 \,r_{\rm g}$ for all the models.
    \label{tab1}}
    \begin{tabular}{
        >{\raggedright}p{0.27\columnwidth} 
	| >{\centering}p{0.15\columnwidth}
        >{\centering}p{0.150\columnwidth}
	>{\centering}p{0.150\columnwidth}
        >{\centering\arraybackslash}p{0.15\columnwidth}
    }
    \hline%\hline
        Model & ~~$V_0/c$~~ & ~~$H/r_{\rm g}$~~ &~~$H/r_{\rm a}$~~ &~~$z_{\rm cut}/H$~~\\
        \hline %\hline
    \texttt{M032.1000.10} & 0.032 &1000 & 0.512 &10 \\
    \texttt{M040.1000.10} & 0.04 & 1000 & 0.800  &10 \\
    \texttt{M050.1000.10} & 0.05 & 1000 & 1.25  &10 \\
    \texttt{M060.1000.10} & 0.06 & 1000 & 1.8 & 10 \\
    \texttt{M080.1000.10} & 0.08 & 1000 & 3.20 &10 \\
    \texttt{M100.1000.10}  & 0.1  & 1000 & 5    &10 \\
    \texttt{M200.1000.10}  & 0.2  & 1000 & 20   &10 \\
    \texttt{M050.500.10}  & 0.05 & 500  & 0.625  &10 \\
    \texttt{M100.500.10}   & 0.1  & 500  & 2.50  &10 \\
    \texttt{M200.500.10}   & 0.2  & 500  & 10   &10 \\
    \texttt{M050.200.10}  & 0.05 & 200  & 0.250 &10 \\
    \texttt{M100.200.10}   & 0.1  & 200  & 1    &10 \\
    \texttt{M200.200.10}   & 0.2  & 200  & 1    &10 \\
    \texttt{M050.1000.05}  & 0.05 & 1000 & 1.25  & 5 \\
    \texttt{M050.1000.20}  & 0.05 & 1000 & 1.25  &20 \\
    \texttt{M050.1000.40} & 0.05 & 1000 & 1.25  &40 \\
    \texttt{M100.1000.20}  & 0.1  & 1000 & 5    &20 \\
    \texttt{M100.1000.40}  & 0.1  & 1000 & 5    &40 \\
    \texttt{M200.1000.20}  & 0.2  & 1000 & 20    &20 \\
    \texttt{M200.1000.40}  & 0.2  & 1000 & 20    &40 \\
        \hline 
\end{tabular}
\end{table}

\begin{figure*}[ht]
\includegraphics[width=\textwidth]{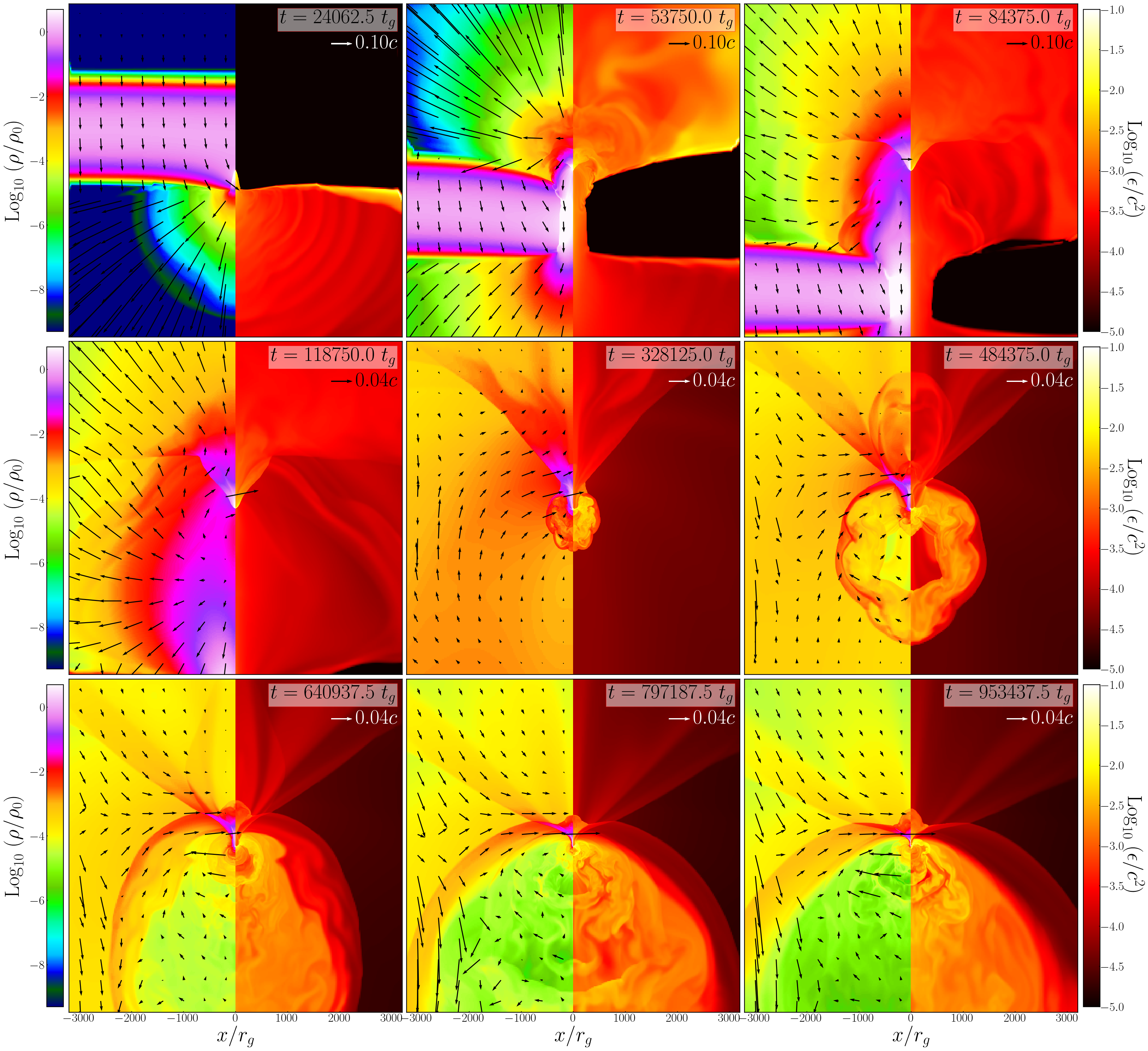}
\caption{Snapshots of the profiles of the rest-mass density and specific internal energy at selected time slices for model \texttt{M050.1000.10}.
For each snapshot, the left and right show the profiles of the rest-mass density in units of the initial maximum density $\rho_0$ and specific internal energy in units of $c^2$, respectively.
Time is shown in units of $t_{\rm g}$. The arrows represent the velocity field of $u^i/u^t$.
Animation of model \texttt{M050.1000.10} is found on \href{https://www2.yukawa.kyoto-u.ac.jp/~masaru.shibata/M050.1000.10.mp4}
{https://www2.yukawa.kyoto-u.ac.jp/$\sim$masaru.shibata/M050.1000.10.mp4}.
}
\label{fig1}
\end{figure*}

\begin{figure*}[ht]
\includegraphics[width=\textwidth]{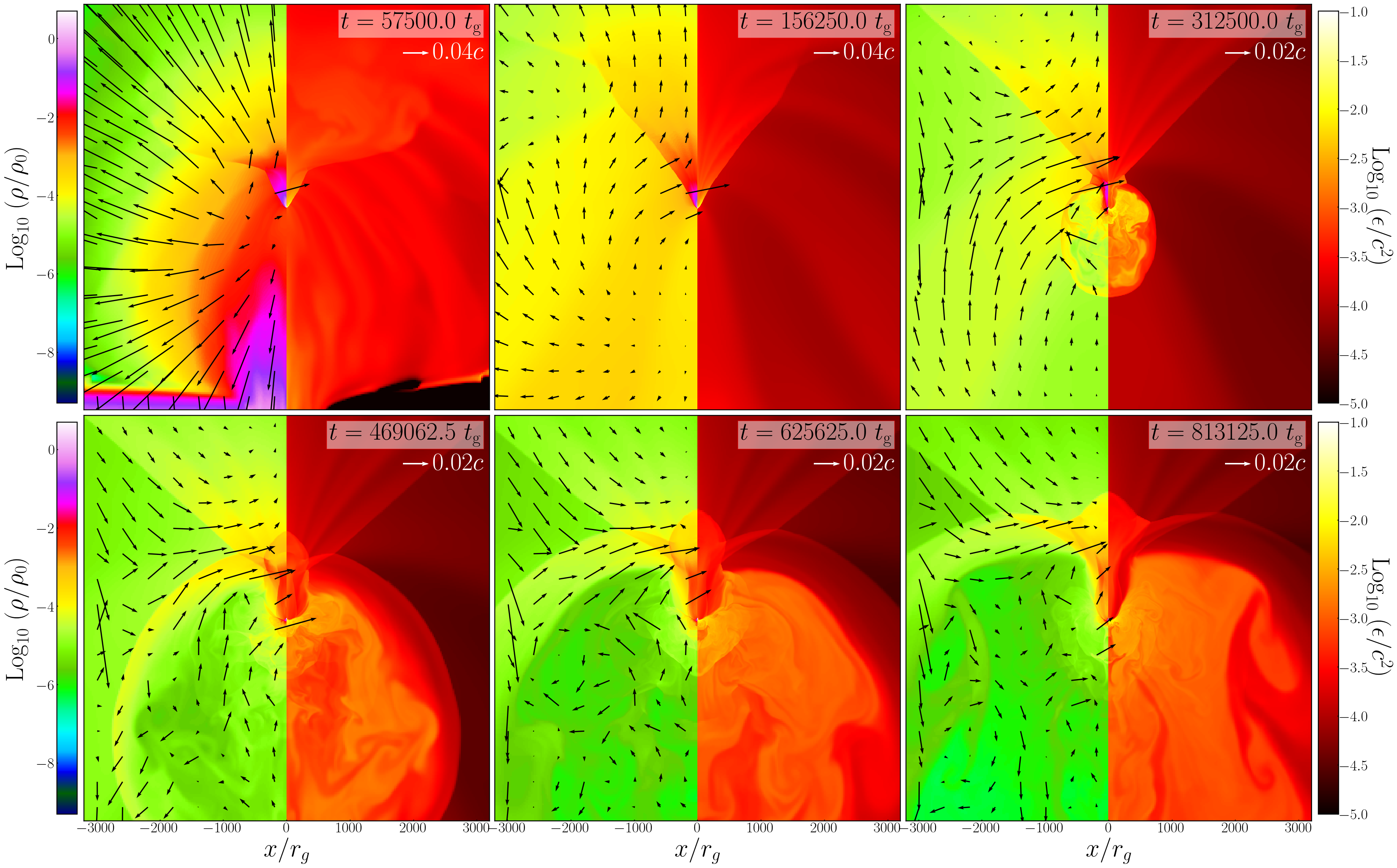}
\caption{The same as \cref{fig1} but for model \texttt{M100.1000.10} at late times after the black hole passes through the disk. 
Animation of model \texttt{M100.1000.10} is found on \href{https://www2.yukawa.kyoto-u.ac.jp/~masaru.shibata/M100.1000.10.mp4}
{https://www2.yukawa.kyoto-u.ac.jp/$\sim$masaru.shibata/M100.1000.10.mp4}.
}
\label{fig1a}
\end{figure*}

\begin{figure*}[ht]
\includegraphics[width=\textwidth]{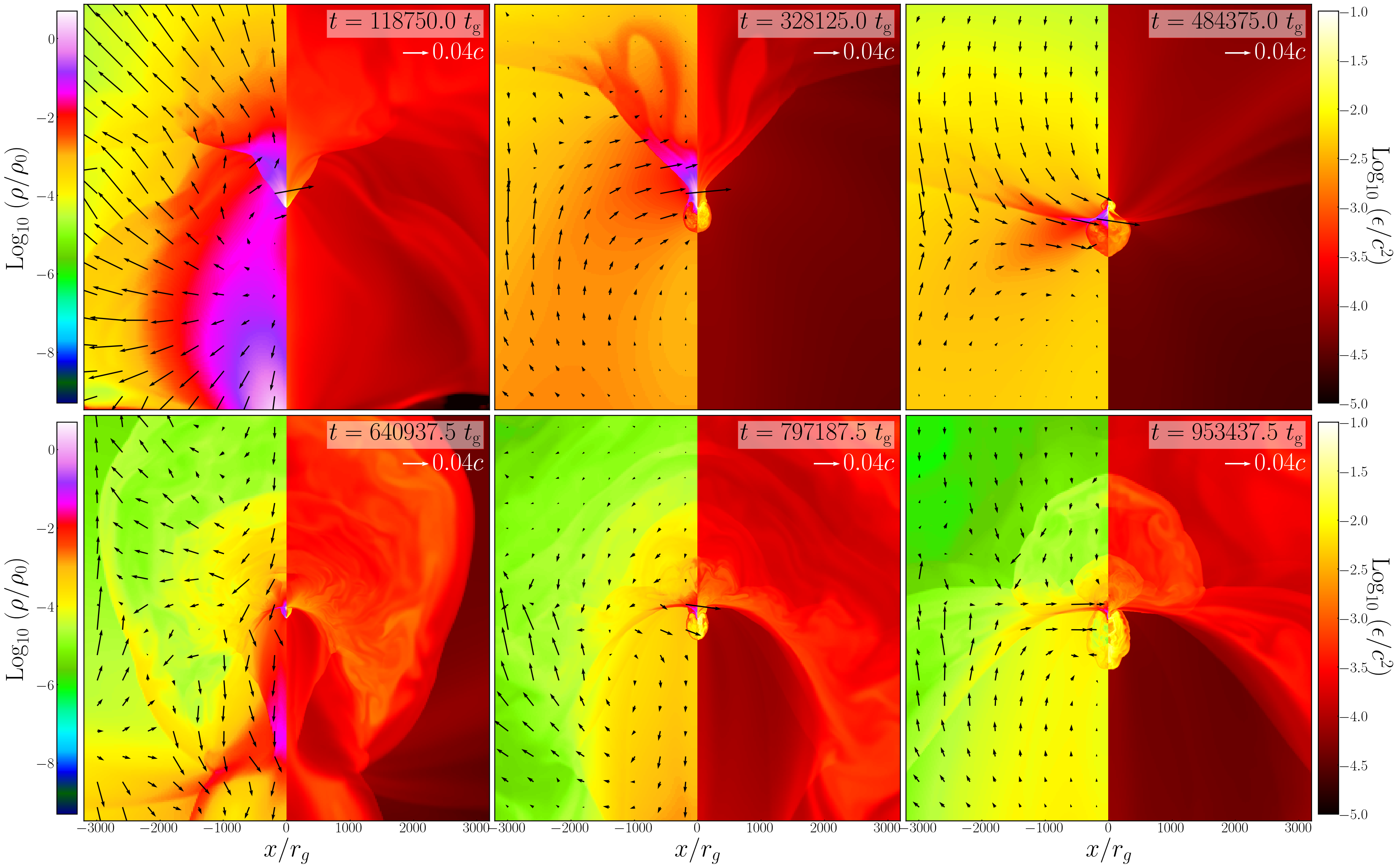}
\caption{The same as \cref{fig1} but for model \texttt{M050.1000.20}. The chosen time slices are the same as for the last 6 snapshots of \cref{fig1}.
Animation of model \texttt{M050.1000.20} is found on \href{https://www2.yukawa.kyoto-u.ac.jp/~masaru.shibata/M050.1000.20.mp4} 
{https://www2.yukawa.kyoto-u.ac.jp/$\sim$masaru.shibata/M050.1000.20.mp4}.
}
\label{fig1b}
\end{figure*}

We perform simulations across a range of velocities $V_0/c=0.032$--$0.2$ and for various disk thickness $H/r_{\rm g}=1000$, 500, and 200 ($H/r_{\rm a}=1/4$--$20$), using $H/r_{\rm g}=1000$ as the fiducial value (see all employed parameters in \cref{tab1}).
Here, the ratio between the disk thickness and the orbital radius $H/R$ is written as
\begin{align}
\begin{split}
\frac{H}{R} &= 0.025 \left(\frac{H}{1000r_{\rm g}}\right)
\left(\frac{Q}{100}\right)^{-1} v_{0.05}^2.
\end{split}
\end{align}
Thus, by comparing it with \cref{eq16} we can identify the corresponding value of $Q$ as
\begin{equation}
Q \approx 25 \left(\frac{c_{\rm s,max}}{10^{-3}c}\right)^{-1}
\left(\frac{H}{1000\,r_{\rm g}}\right) v_{0.05}^3.
\end{equation}
The gravity cutoff expressed as $z_{\rm cut}/H$ is also varied in the range 5--40 as previously mentioned in \cref{sec2}.
In the end, we find that the results depend only weakly on disk thickness $H/r_{\rm g}$ in this chosen range, while the dependence of the mass accretion rate onto the black hole and energy distribution for the outflow matter on velocity $V_0/c$ and the gravity cutoff $z_{\rm cut}$ is significant.

\subsection{General picture of the collision process}

Figure~\ref{fig1} displays the snapshots of the rest-mass density profiles (left at each time) and specific internal energy profiles (right) at selected time slices for model \texttt{M050.1000.10}.
Irrespective of the chosen parameters, the evolution process in the early stage is qualitatively very similar to each other.
As already reported by the authors of \cite{1998ApJ...507..131I}, two outflows are driven from the collision toward parallel and anti-parallel directions of the relative motion (see the first and second panels of \cref{fig1}).
In the following, we refer to these outflows as the first (downward) and second (upward) components.
Since the properties of these two outflows are quantitatively different, we analyze them separately in the following. 

\begin{figure*}[t]
\includegraphics[width=0.98\textwidth]{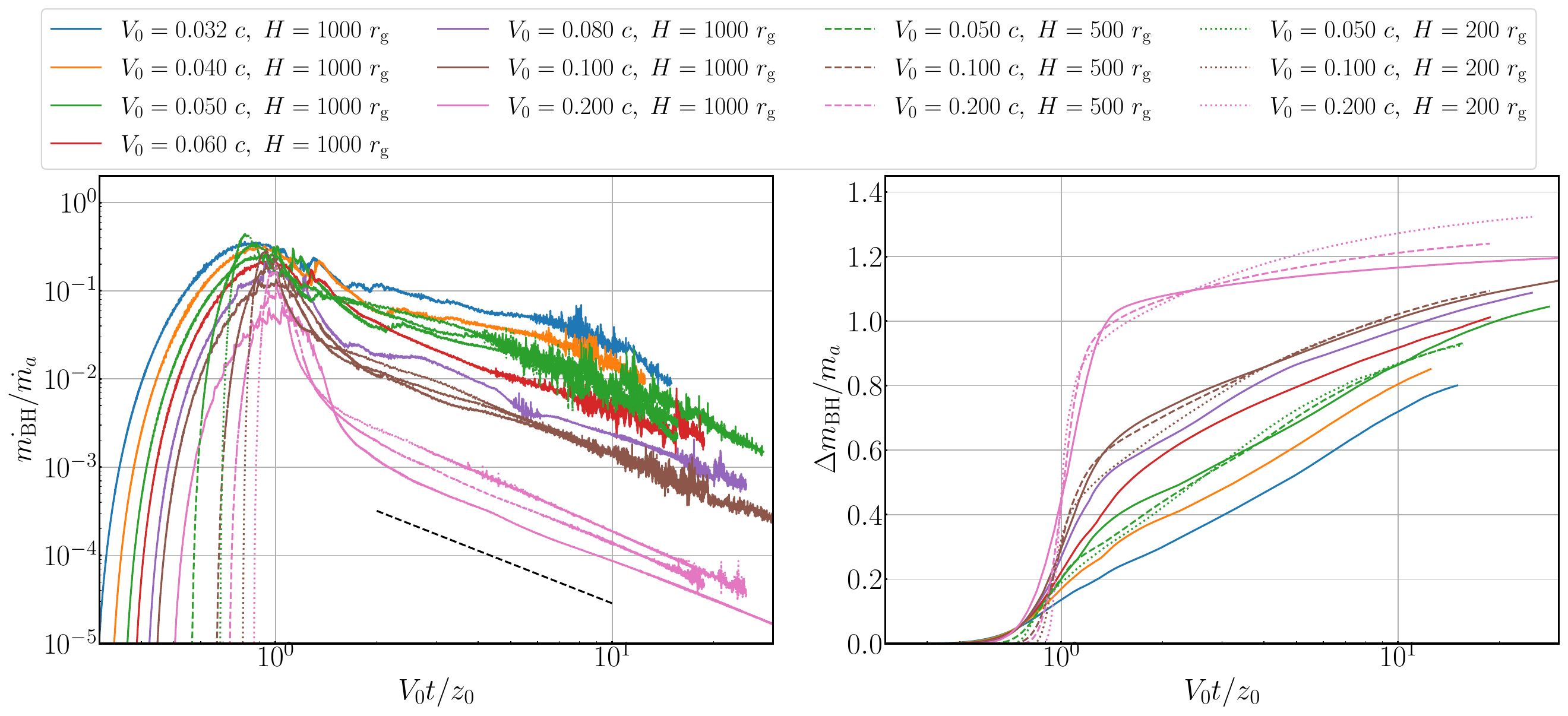}\\
\includegraphics[width=0.98\textwidth]{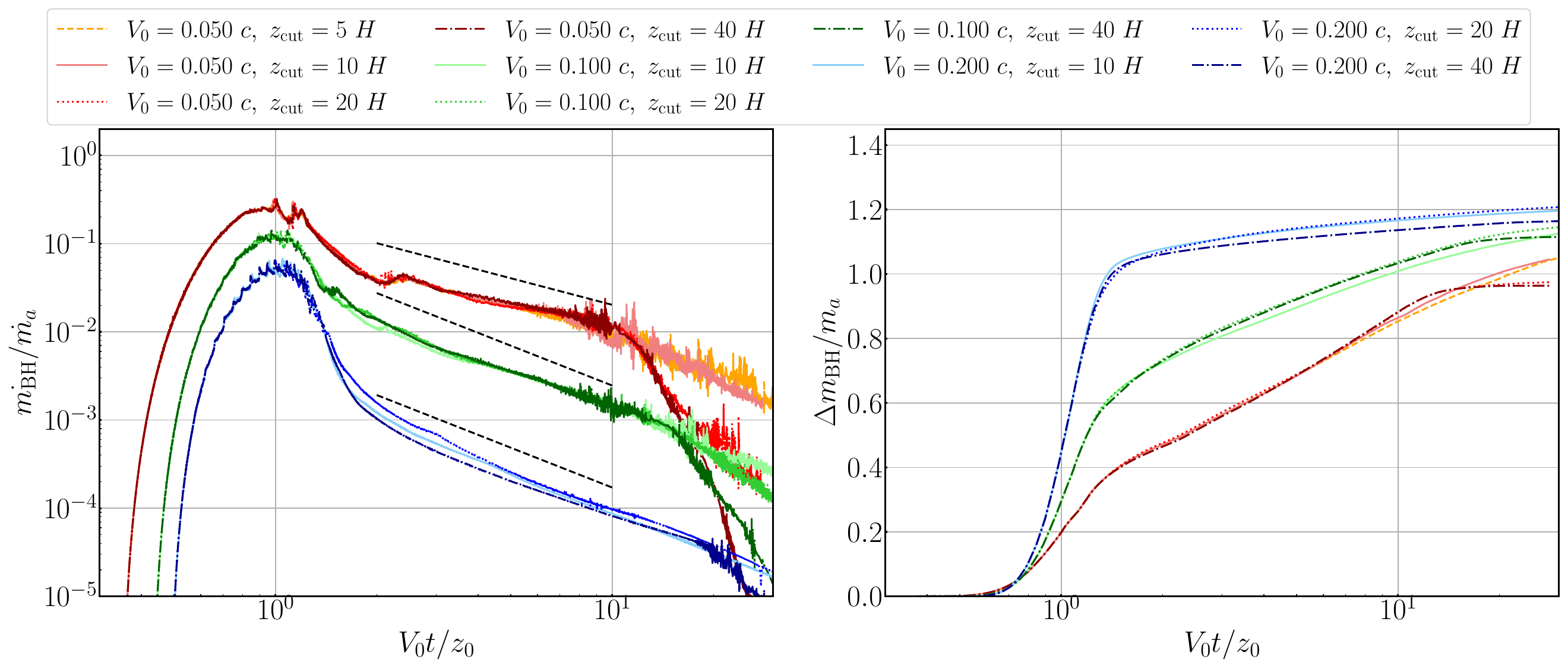}
\caption{Upper left: Evolution of $\dot m_{\rm BH}$ for models with $z_{\rm cut}=10H$ but for a variety of $V_0/c$ and $H/r_{\rm g}$. Upper right: The same as the upper left panel, but for the evolution of $m_{\rm BH}$.
Lower left and right: The same as upper left and right, respectively, but for a variety of $z_{\rm cut}/H$ with $V_0/c=0.05$, 0.1, 0.2, and $H=1000\,r_{\rm g}$. 
For all the panels, the time is shown in units of $z_0/V_0$.
The dashed slopes in the left panels represent $\propto t^{-n}$ with $n=1.5$ in the upper panel, and $n=1$, 1.5, and 1.5 in the bottom panel, which are plotted as counterparts of the models with $V_0/c=0.05$, 0.1, and 0.2, respectively.
The oscillatory behavior in the late stage of $\dot m_{\rm BH}$ is associated with the presence of convective motion around the black hole (cf. \cref{fig1,fig1a,fig1b}).
}
\label{fig2}
\end{figure*}

After the black hole passes through the disk, a cylindrical high-density region with a radius of $\sim 400 \,r_{\rm g}\sim r_{\rm a}/2$ is formed in the disk (see the second and third panels of \cref{fig1}) and then gradually expands.
In addition to the outflow, bumps with a height a few times larger than $H$ are formed above the disk on both sides (see the third panel of \cref{fig1}).
On the upper side, the accretion onto the black hole continues from the bump for a long timescale (see the last 6 panels of \cref{fig1} and \cref{fig2}), and a quasi-steady accretion shock with a cone shape is generated around the black hole (see the third and fourth panels of \cref{fig1}).
Inside the accretion shock, a high-density and high-temperature region is formed, and from there, matter accretion continues for a long timescale (see \cref{sec:bh_acc}).
All these findings agree qualitatively with those in \cite{1998ApJ...507..131I}.

One new interesting finding is that in the late-stage evolution, convection occurs upstream of the accretion shock (see the last 5 panels of \cref{fig1}).
As a result, the high-internal-energy region expands with the radius of $\agt 1000\,r_{\rm g}$.
The activity of this convection zone depends strongly on $V_0/c$ and $z_{\rm cut}$ (see \cref{sec:bh_acc}).

The evolution process described above does not strongly depend on $H$ for a given value of $V_0$.
On the other hand, the evolution process after the black hole passes through the disk depends quantitatively on the choice of $V_0$ due to its significant dependence on the accretion radius $r_{\rm a}$.
For large values of $V_0/c \agt 0.1$, the mass inflow rate toward the black hole swiftly shuts down after the black hole passes through the disk (cf.~\cref{fig2}).
By contrast, for small values of $V_0/c \alt 0.1$, the high-mass inflow rate continues for a long timescale $\sim 10^6 \,t_{\rm g}$, reflecting that the matter falls toward the black hole from a wide region of the disk with $\alt 2r_{\rm a}$ (note that for model \texttt{M050.1000.10}, $r_{\rm a}=800\,r_{\rm g}$).
The fallback from the pre-ejected matter also plays an important role for a large value of $z_{\rm cut}$ (see \cref{sec:bh_acc}). 

Figure~\ref{fig1a} displays the late-time profiles of the rest-mass density and specific internal energy for model \texttt{M100.1000.10} for the comparison with those of model \texttt{M050.1000.10} shown in \cref{fig1}.
It is found that the density around the black hole for \texttt{M050.1000.10} is much higher than that for \texttt{M100.1000.10} until the late time of the evolution, indicating that a large mass inflow to the black hole continues for a longer timescale.
Associated with the high-mass accretion toward the accretion shock, thermal energy near the shock is enhanced for \texttt{M050.1000.10}, while for \texttt{M100.1000.10}, such a high thermal-energy region is located only near the black hole.
For \texttt{M100.1000.10}, convection is enhanced in a late stage as in \texttt{M050.1000.10}, but the activity is weaker than for \texttt{M050.1000.10}. 

The late-time evolution depends strongly on the choice of $z_{\rm cut}$.
Figure~\ref{fig1b} displays the evolution of the rest-mass density and specific internal energy for model \texttt{M050.1000.20}.
The chosen time slices are the same as the last six snapshots of \cref{fig1}.
In this model, the larger value of $z_{\rm cut}$, which is twice that of model \texttt{M050.1000.10}, results in a stronger gravitational effect from the primary black hole, leading to a more pronounced fall-back toward the disk.
As a result, for $t\agt 4\times 10^5 \,t_{\rm g}$ the matter with $v^z < 0$ dominates the vicinity of the black hole, and the cone-shaped accretion shock alters its morphology as seen in the third panel of \cref{fig1b}, triggering convection for the $z>0$ region.
The convection is subsequently excited more vigorously until a new cone-shaped accretion structure is developed for $t \agt 10^6\,t_{\rm g}$.
An appreciable fraction of the matter goes down toward the disk and hence the mass accretion rate onto the black hole drops in the late stage (cf. \cref{fig2}). 

This late-time fallback is more relevant for smaller values of $V_0/c$. For instance, for $V_0/c=0.032$, the fall-back behavior is found for $t \agt 7 \times 10^5\,t_{\rm g}$ even for $z_{\rm cut}=10H$ (cf.~\href{https://www2.yukawa.kyoto-u.ac.jp/~masaru.shibata/M032.1000.10.mp4} 
{https://www2.yukawa.kyoto-u.ac.jp/$\sim$masaru.shibata/M032.1000.10.mp4}).
Notably, as the fallback process ends and the density of the infalling matter rapidly decreases, the hot matter surrounding the black hole explodes towards the positive $z$-direction, a phenomenon reminiscent of the shock breakout that occurs during the early stages of a supernova explosion.

\begin{figure*}[t]
\includegraphics[width=0.498\textwidth]{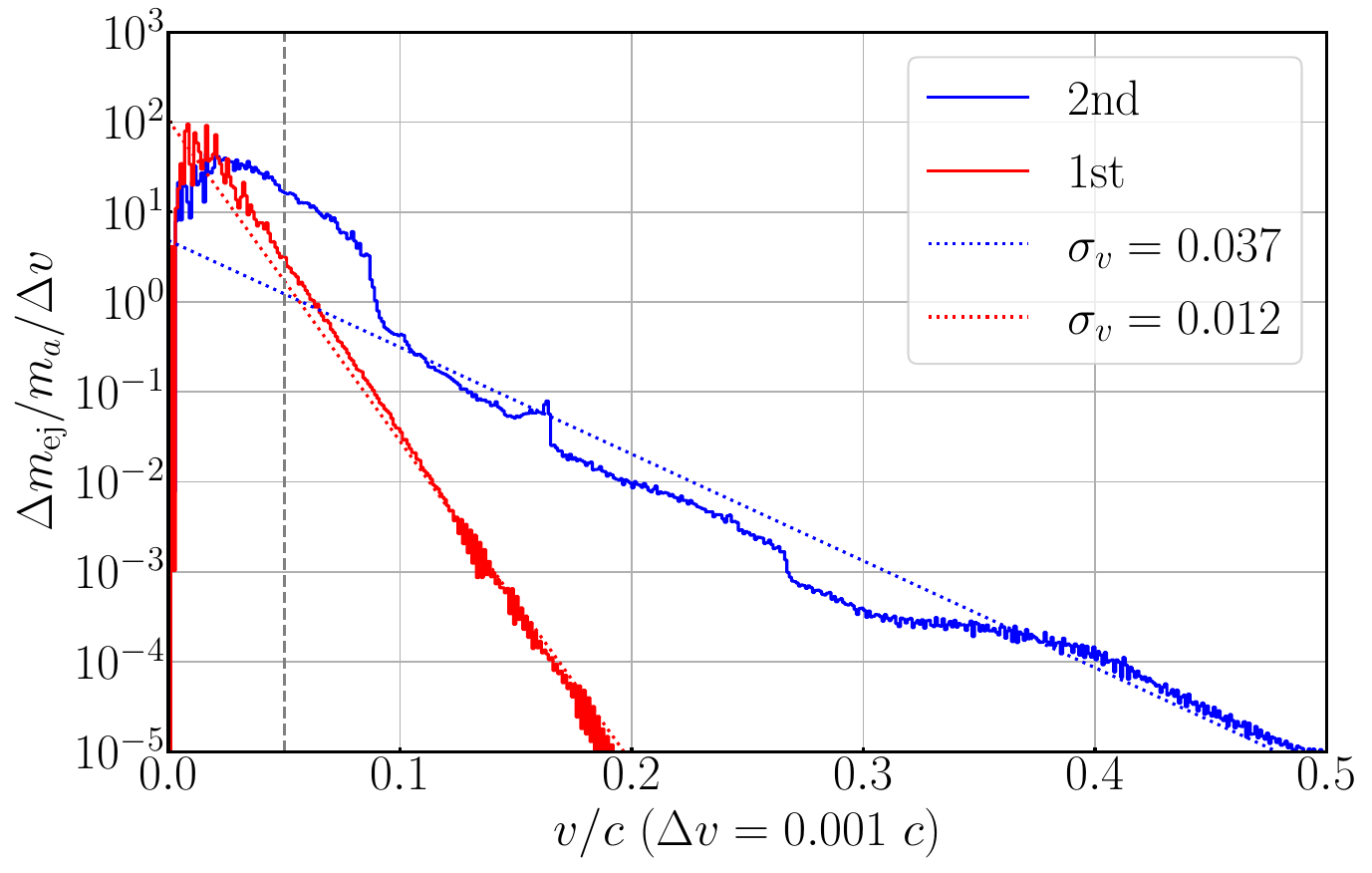}
\hspace{-5mm}
\includegraphics[width=0.498\textwidth]{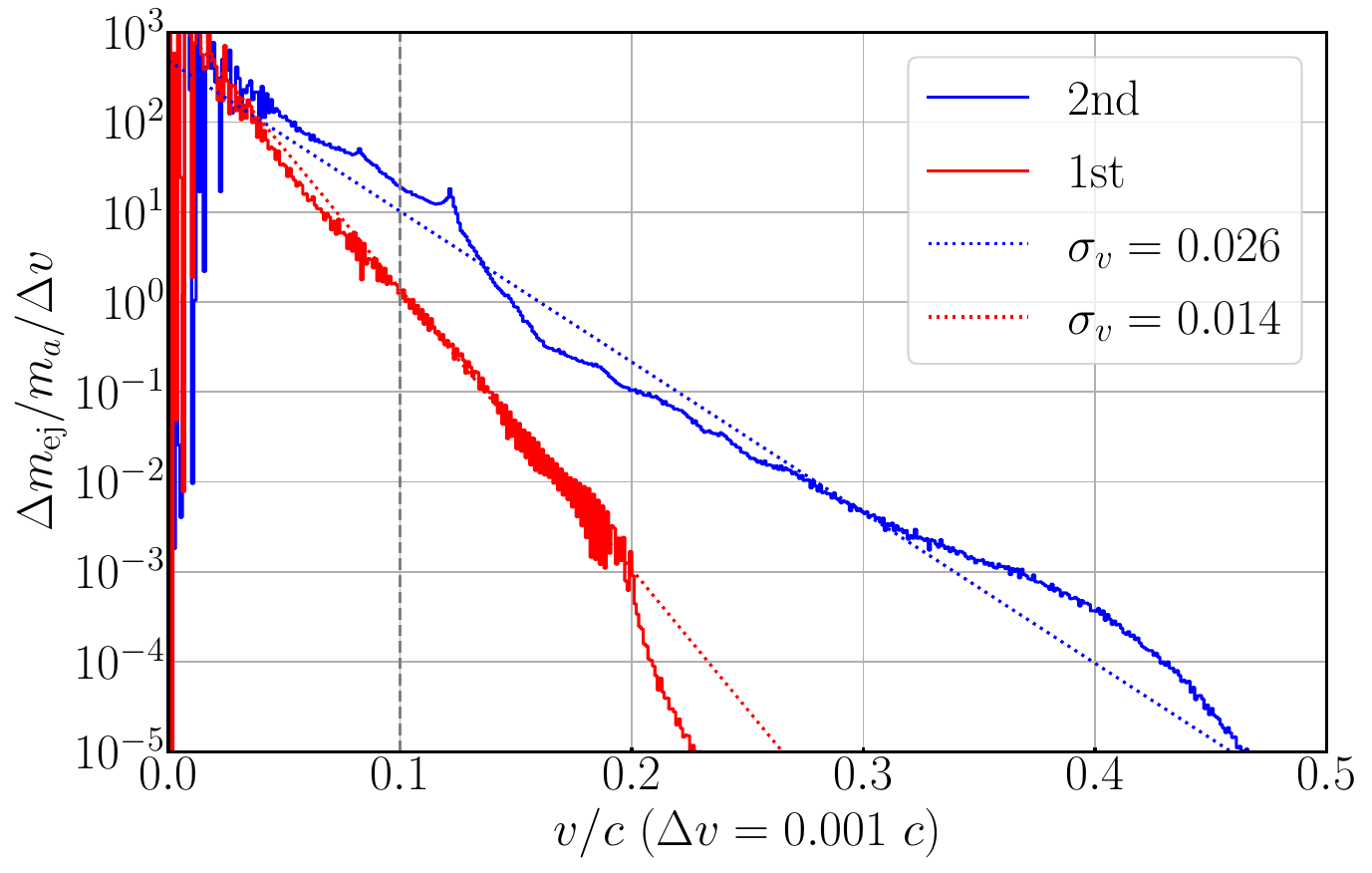}\\
\includegraphics[width=0.498\textwidth]{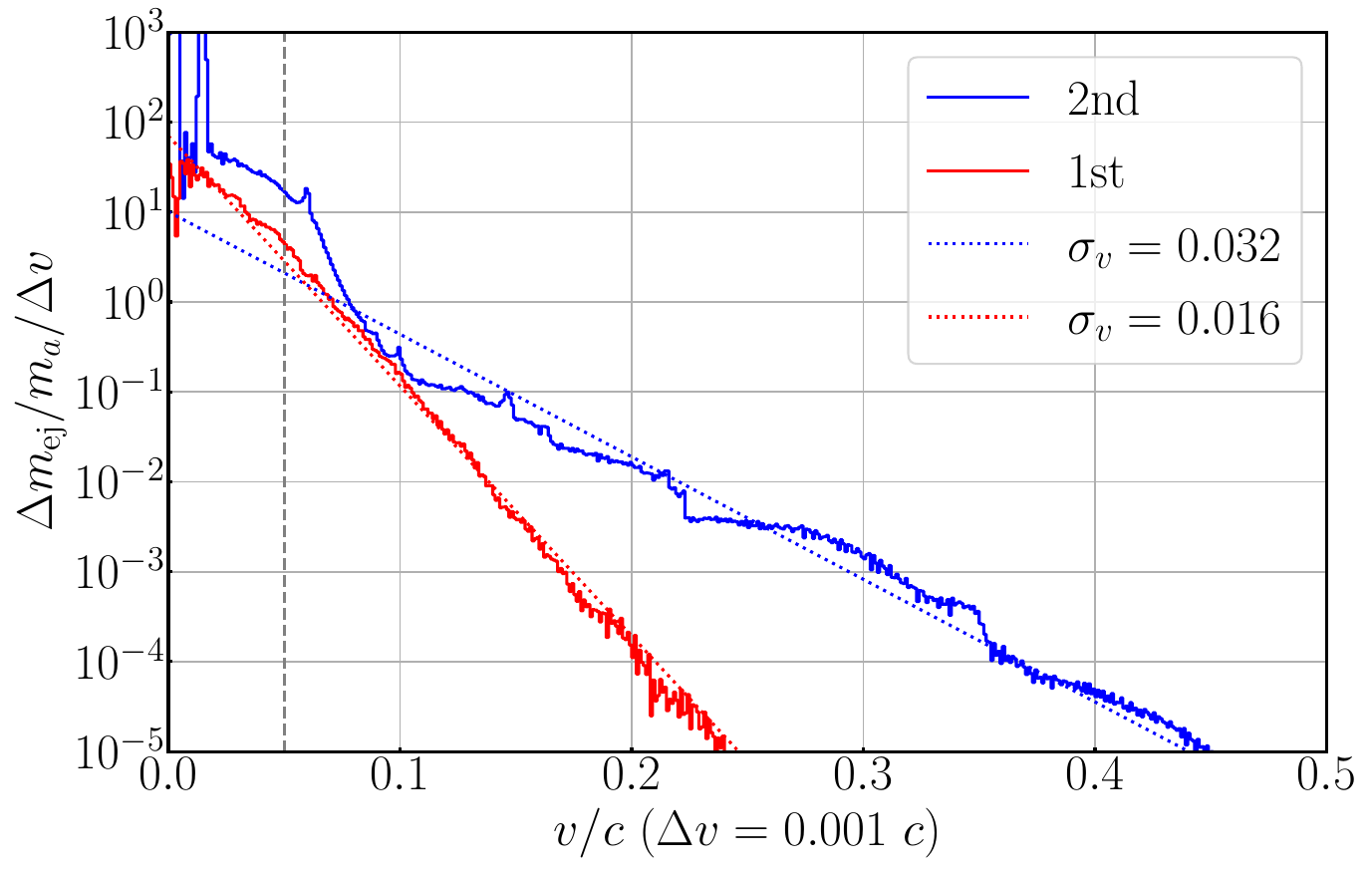}
\hspace{-5mm}
\includegraphics[width=0.498\textwidth]{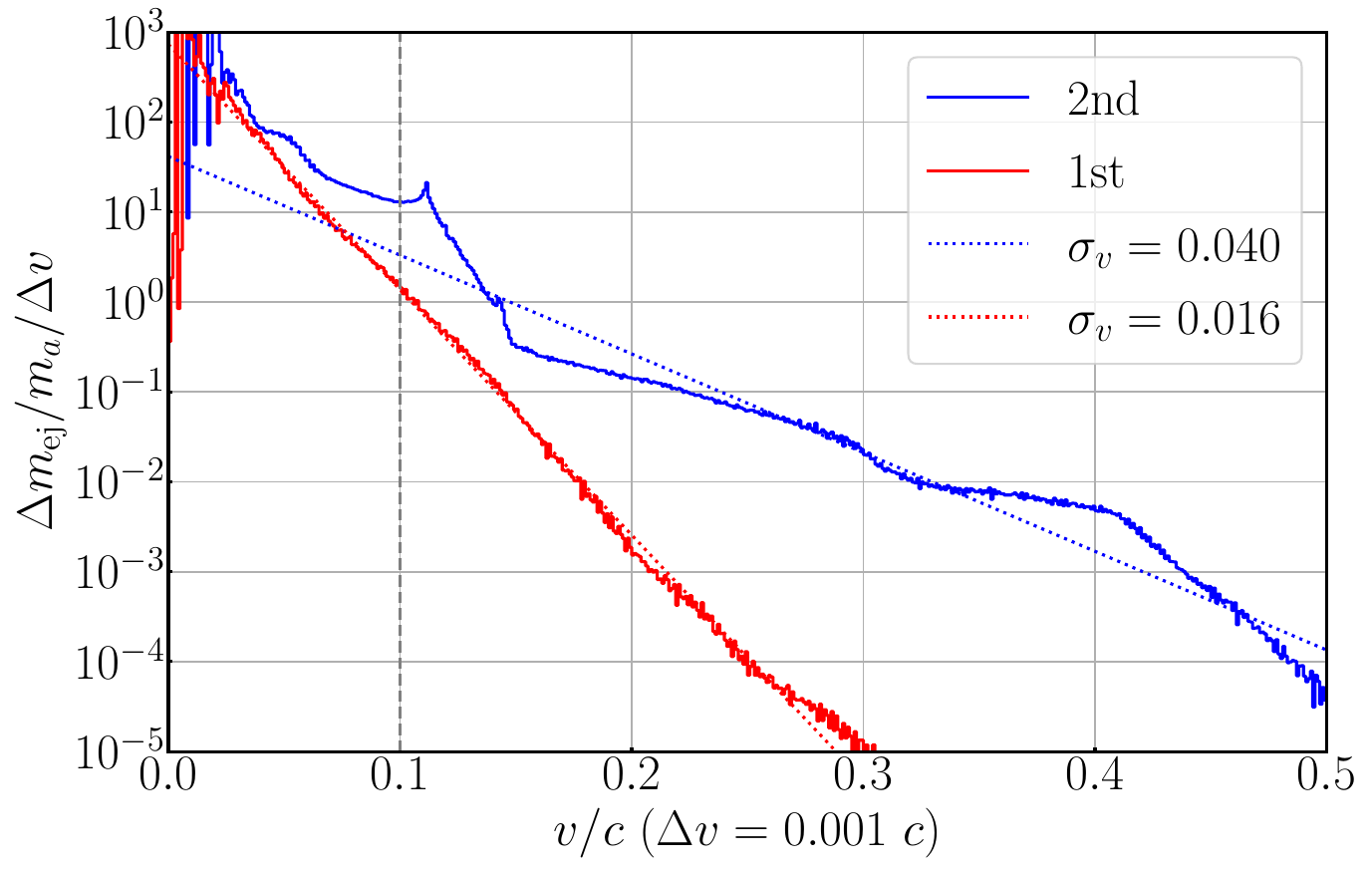}
\caption{$\Delta m_{\rm ej}$ in the velocity bin of $\Delta v=0.001c$ in units of $m_{\rm a}$ for models \texttt{M050.1000.10} (upper left), \texttt{M100.1000.10} (upper right), \texttt{M050.200.10} (lower left), and \texttt{M100.200.10} (lower right) at $t=10^5\,t_{\rm g}$.
The grey vertical line denotes $v=V_0$, and we pay attention only to the components of $v > V_0$. 
}
\label{fig5}
\end{figure*}

\subsection{Mass accretion on the black hole} \label{sec:bh_acc}

Figure~\ref{fig2} shows the evolution of mass accretion rate onto the black hole $\dot m_{\rm BH}$ (left) and the total accreted rest mass $\Delta m_{\rm BH}$ (right) as functions of time in units of $\dot m_{\rm a}$ and $m_{\rm a}$, respectively.
The upper panels compare the results for the models with $z_{\rm cut}=10H$ but for a variety of disk velocity $V_0/c\,(=0.032$--$0.20)$ and thickness $H/r_{\rm g}\,(=200$--$1000)$. 
The collision between the black hole and the disk sets in at a time of $\alt (z_0-H)/V_0$, at which $\dot m_{\rm BH}$ steeply increases irrespective of the parameters $(V_0, H)$.
The peak accretion rate onto the black hole $\dot m_{\rm BH}$ is reached approximately at the time $t \sim z_0/V_0$, although this peak time is slightly earlier for smaller initial velocities $V_0$.
The peak value of the accretion rate $\dot m_{\rm BH}/\dot m_{\rm a}$ is broadly of order $\sim 0.1$, and it decreases monotonically with increasing $V_0/c$.
The maximum for the peak value of $\dot m_{\rm BH}$ with respect to $V_0$ is likely to be $\sim 0.3\dot m_{\rm a}$ because of its weak dependence on $V_0/c$ for $V_0/c \leq 0.05$. 

For $V_0/c \geq 0.05$, $\dot m_{\rm BH}$ decreases approximately monotonically after the peak is reached, but the decrease rate gradually becomes gentle, settling to $\propto t^{n}$ where, for example, $n \approx -1$, $-1.5$, and $-1.5$ for $V_0/c=0.05$, 0.1, and 0.2, respectively.
For $0.05 \leq V_0/c \leq 0.1$, the power-law index $n$ lies between $-1.5$ and $-1$.
The gradual decline is due to the ongoing supply of matter from the disk, which continues to feed the black hole even after it has passed through the disk.
The oscillatory behavior observed in the late-stage evolution of $\dot m_{\rm BH}$ for models with $V_0/c \leq 0.1$ is a consequence of the increased convective activity around the black hole.

For $V_0/c \leq 0.05$, the decline of $\dot m_{\rm BH}$ after the peak is reached is gentler than $t^{-1}$, persisting until $V_0t/z_0 \sim 10$.
However, a sharp decline is observed for $V_0 t/z_0 \agt 10$.
This suggests that in the case of a relatively slow collision, the black hole continues to accrete matter at a high rate for a long timescale $\geq 10r_{\rm a}/V_0$ (notably for $V_0/c=0.032$, $r_a \approx z_0$). 

For smaller values of $H/r_{\rm g}$, the peak value of $\dot m_{\rm BH}/\dot m_{\rm a}$ is increased, but the duration of the peak is naturally shorter. Otherwise, we do not observe a significant dependence of $\dot m_{\rm BH}/\dot m_{\rm a}$ on $H$. 

The lower left panel of \cref{fig2} illustrates that for $V_0/c=0.05$, $\dot m_{\rm BH}$ drops steeply for $t \agt 10z_0/V_0$ with $z_{\rm cut}/H \geq 20$.
This implies the fact that the attractive gravity toward the $-z$ direction plays an important role in capturing the matter toward the disk.
A similar effect is observed in the late stages of evolution for $V_0/c \geq 0.1$, where $\dot m_{\rm BH}$ declines steeply for $t \agt 20z_0/V_0$.

The observed power-law index of $n=-1.5$ for $V_0/c \agt 0.1$ may be interpreted by the following analysis.
In the late stage of the evolution, the mass accretion rate onto the black hole can be approximated as $\dot m_{\rm BH} \sim \pi \rho \tilde r_{\rm a}^2 \tilde V$ where $\tilde V$ is the velocity at which matter is injected from the disk, and $\tilde r_{\rm a} \sim 2GM_{\rm BH}/\tilde V^2$.
For simplicity, we neglected the effects of thermal contribution.
Assuming that the density of the accreting material is proportional to $l^{-3}$ where $l$ is the distance between the black hole and the disk, i.e., $l \sim V_0t$, we find $\rho \propto t^{-3}$ with $t$ being the time measured from the moment that the black hole has passed through the disk.
$\tilde V$ may be broadly estimated by $\sqrt{GM_{\rm BH}/l} \propto t^{-1/2}$. 
Therefore, we can obtain $\dot m_{\rm BH} \propto t^{-3/2}$ with this simple interpretation.

However, this analysis does not account for the shallower slope observed for $t \alt 10 z_0/V_0$ with $V_0/c \alt 0.05$.
The slower decline of $\dot m_{\rm BH}$ for the smaller values of $V_0$ is likely due to the fact that the black hole remains close to the disk for a longer timescale and $r_{\rm a} \propto V_0^{-2}$ significantly enhances the outflow from a wider region of the disk.
As a result, the matter ejected from the disk is distributed over a larger region around the black hole for smaller values of $V_0/c$,
leading to accretion from both the $z$ direction and the $x$ direction.
This appears to enhance the matter accretion for the models with the smaller values of $V_0/c$.

The right panels of \cref{fig2} show that the values of $\Delta m_{\rm BH}$ approach $\sim m_{\rm a}$ irrespective of the model parameters.
However, the manner in which $\Delta m_{\rm BH}$ approaches $m_{\rm a}$ depends on $V_0/c$, which is consistent with our earlier findings regarding the evolution of $\dot m_{\rm BH}$.
On the other hand, the dependence of the $\Delta m_{\rm BH}$ evolution path on $H$ is relatively weak, with only a slight increase in $\Delta m_{\rm BH}/m_{\rm a}$ observed as $H$ increase.
This increase is attributed to the widening of the cylindrical shocked region inside the disk as the black hole spends more time within the disk for larger values of $H$.
Nonetheless, this effect is relatively minor.
The dependence of $m_{\rm a}$ curve on $z_{\rm cut}$ is only clearly seen in the late time with $V_0/c \leq 0.05$. 

One noteworthy point is that for $V_0/c \alt 0.1$, the accretion rate $\dot m_{\rm BH}$ remains high for a long timescale.
As we showed in \cref{sec2}, $\dot m_{\rm BH}$ can significantly exceed the Eddington accretion rate $\dot m_{\rm Edd}$, if the column density of the disk $\Sigma$ is sufficiently (but reasonably) high, implying that the mass accretion rate can be higher than $\dot m_{\rm Edd}$ for a long timescale of $\sim 10^6\,t_{\rm g} \approx 14 (M_{\rm BH}/10^4M_\odot)\,{\rm h}$.
Assuming circular orbits, the orbital period of the black hole is written as $P_\mathrm{orb}\approx 2.9\, v_{0.05}^{-3} (M_{\rm c}/10^6M_\odot) \,{\rm d}$.
Thus, the duration of the super-Eddington accretion phase becomes a substantial fraction of the orbital period for the case of $Q=O(10)$, although for $Q\sim 100$ the ratio of $10^6t_\mathrm{g}/P_\mathrm{orb}$ is comparable to the ratio of the duration to the recurrence time of typical QPEs. 
The implications for this will be explored in \cref{sec5}.

\begin{figure*}[t]
\includegraphics[width=0.98\textwidth]{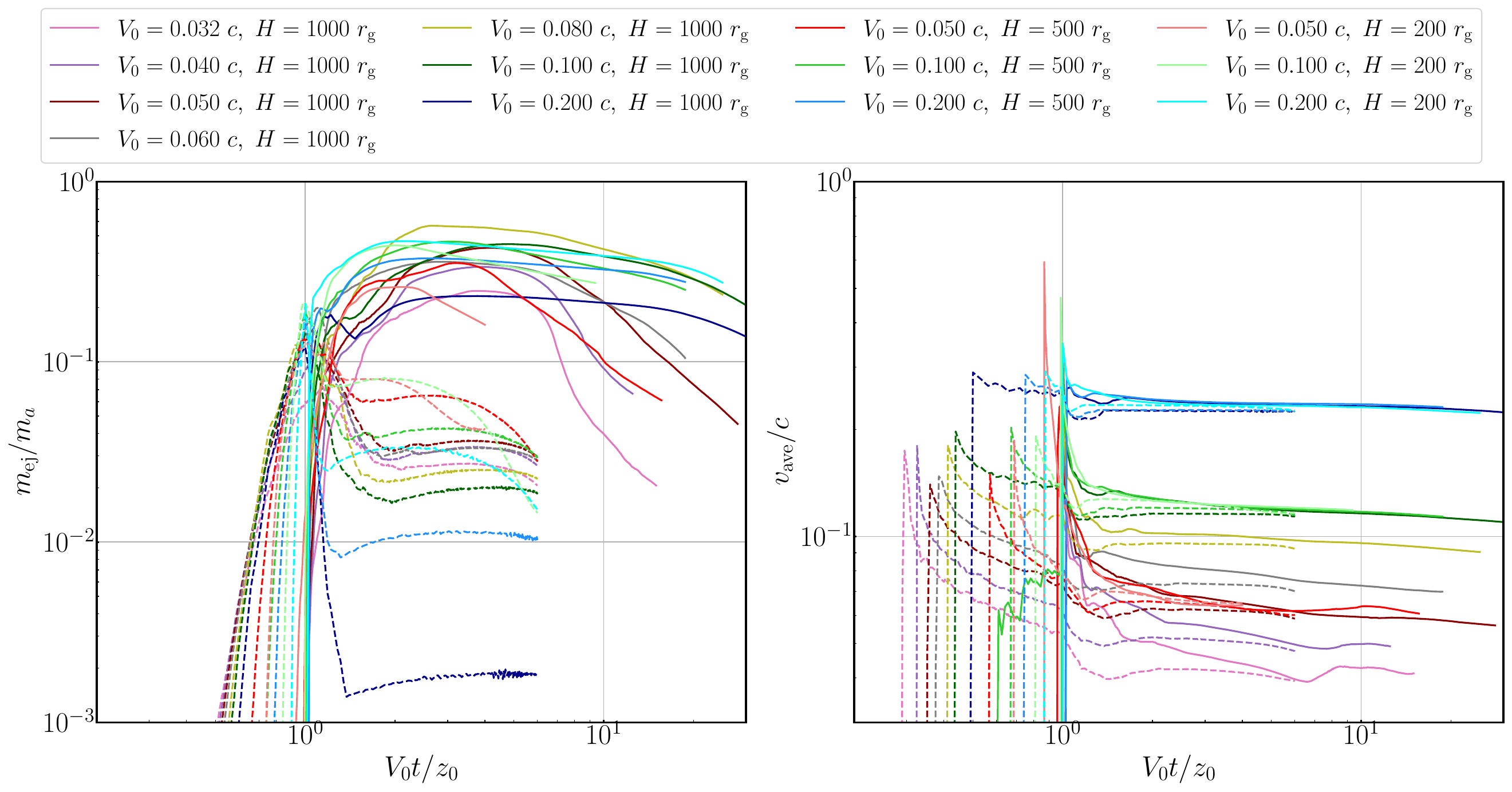}
\vspace{-4mm}
\caption{Left: Evolution of $m_{\rm ej}$ in units of $m_{\rm a}$ as a function of time for a wide variety of models.
The dashed and solid curves represent the ejecta mass for the first and second outflows, respectively.
The decrease in $m_{\rm ej}$ in the late stage is attributed to two factors: 
(i) a portion of the matter escapes from the computational domain, and (ii) a portion of the matter is captured by the hypothetical gravity from the central massive black hole $M_c$, effectively preventing it from being ejected.
Right: The same as the left panel, but for the average velocity. 
}
\label{fig6}
\end{figure*}

\subsection{Properties of outflow} \label{sec:outflow}

Figure~\ref{fig5} shows the outflow mass spectrum $dm_{\rm ej}/dv$ for models \texttt{M050.1000.10} (upper left), \texttt{M100.1000.10} (upper right), \texttt{M050.200.10} (lower left), and \texttt{M100.200.10} (lower right).
For $dm_{\rm ej}/dv$, we plot the ejecta mass $\Delta m_{\rm ej}$ in the velocity bin with a width of $\Delta v = 0.001c$.
The red and blue curves denote the results for the first (downward) and second (upward) outflows, respectively.
We pay attention only to the components of $v >V_0$ as the matter with $v < V_0$ would return to the disk or fall into the primary black hole in reality.
For the first outflow, the majority has the velocity of $v \sim V_0$ as the outflow mass spectrum $dm_{\rm ej}/dv$ decreases steeply with increasing velocity $v$. 
On the other hand, the second outflow has a higher-velocity component that behaves approximately like $dm_{\rm ej}/dv \propto \exp(-v/\sigma_v)$, with $\sigma_v/c \sim 0.03$--0.04 as shown as the dotted slopes in \cref{fig5}.
When the black hole passes through the disk, these high-velocity components are accelerated by the gravity of the passing-through black hole as well as through the steep density gradient near the disk's surface, and ejected into the nearly vacuum region in earlier times. 
Although the high-velocity components are present, the majority of the outflow has a velocity of $v \alt 2V_0$, and thus, the total kinetic energy is of order $m_{\rm ej}V_0^2$.
This trend is observed in all models as shown in \cref{fig5}, regardless of the disk thickness $H$.
By analyzing $dm_{\rm ej}/dv$ for other values of $V_0$, we also found similar results for $0.032 \alt V_0/c \alt 0.2$.

We find that the second outflow ejects more mass due to a significant component at a velocity $v \sim V_0$, but the total ejecta mass is of order $\sim 10\%$ of $m_{\rm a}$.
Figure~\ref{fig6} illustrates the evolution of the ejecta mass $m_{\rm ej}/m_{\rm a}$ as a function of time with the dashed and solid lines representing the first and second outflows, respectively.
Here, the ejecta mass is defined as
\begin{equation}
m_{\rm ej}:=\int_{V_0}^c dv \frac{dm_{\rm ej}}{dv},
\end{equation}
i.e., the integration is performed only for the components of $v \geq V_0$.
The first and second outflows are integrated within the regions below ($z<\bar z$) and above ($z>\bar z$) the disk plane, respectively.
It is found that the ejecta mass $m_{\rm ej}$ is smaller than $m_{\rm a}$ for both outflows, particularly for the first outflow.
This is primarily because most of the initially outflowed matter subsequently returns to the disk by the hypothetical gravitational effect from the primary black hole $M_c$, as described in \cref{eq1}, or is captured by the secondary black hole $M_{\rm BH}$.
This observation is supported by the fact that the ejecta mass drops slightly for larger values of $z_{\rm cut}$ as expected from the bottom panels of \cref{fig2}.
Indeed, if the initial velocity at the disk surface is smaller than $V_0$, the matter should return to the disk because of the gravity from the primary black hole.
Note that a previous study \cite{1998ApJ...507..131I} suggested that the outflow is driven from both sides significantly, but their conclusion was based on a model that did not account for the gravitational effect of the primary black hole.

The right panel of \cref{fig6} shows the average velocity $v_{\rm ave}$ of the first (dashed) and second (solid) outflow defined by
\begin{equation}
v_{\rm ave}:= \left[(m_{\rm ej}+\epsilon m_{\rm a})^{-1}\int_{V_0}^c dv \frac{dm_{\rm ej}}{dv} v^2\right]^{1/2},
\end{equation}
where we set $\epsilon=10^{-12}$ to focus on the cases with the appreciable ejecta mass. 
This clearly illustrates the difference between the first (dashed) and second (solid) outflows.
For the first outflow, the average velocity remains consistently below $2V_0$, with typical values being 20--$30\%$ greater than $V_0$.
In contrast, the second outflow can reach higher velocities, up to $\sim 0.6c$, especially at the onset of mass ejection.
During these early stages, the ejecta mass is still relatively small as shown in the left panel of \cref{fig6}, resulting in a low mass fraction of the high-velocity components observed in \cref{fig5}.
Nevertheless, this figure clearly demonstrates the difference in the mechanisms of the outflow launches between the first and second instances.
In the late stage, the average velocity of the second outflow decreases and approaches that of the first one.
This occurs because the proportion of low-velocity ejecta increases over time and some of the high-velocity matter escapes from the computational domain earlier.
As indicated in \cref{fig5}, the asymptotic average velocity of the second outflow is expected to be 20--30\,\% larger than $V_0$ similar to the first outflow.
Therefore, in general terms, the total ejecta energy for both outflows can be approximated as $\sim m_{\rm ej}V_0^2$ . 

\section{Discussion}\label{sec5}

\subsection{Relation with QPE}\label{sec4.1}

We begin by focusing on the electromagnetic counterparts of the matter ejected when the black hole collides with the disk.
Using a method similar to the one applied in the analysis of supernova explosions~\cite{1998ApJ...507..131I, 1980ApJ...237..541A}, we estimate the peak luminosity emitted by the expanding outflow matter and the time at which this peak occurs, assuming an initial mass $m_{\rm ej}$ and internal energy $E_{\rm ej}$ of the ejected matter.
In the following, we denote the average velocity as $v$, which is approximately equal to or slightly larger than $V_0$, as we have confirmed in this paper.
We analyze the results separately for the first (downward) and second (upward) outflows based on $m_{\rm ej}$ and $E_{\rm ej}$ assuming $E_\mathrm{ej}\sim m_\mathrm{ej}V_0^2$.

For simplicity, we further assume spherical expansion and opacity of $\kappa$ for the outflow matter.
Thus, the diffusion timescale can be approximately expressed as
\begin{align}
t_{\rm diff} \approx \frac{3\kappa m_{\rm ej}}{4\pi r c},
\end{align}
where $r$ is the radial coordinate from the center of the sphere, and we assume that the density decreases steeply with increasing $r$.
Assuming that the gas is composed of ionized hydrogen and helium, the value of $\kappa$ is determined by the Thomson scattering and is set to $\approx 0.35\,{\rm g/cm^2}$.
We define $t_{\rm a}:=r_{\rm a}/v$ which is estimated to be
\begin{align}
t_{\rm a} \approx 790 M_{\rm BH,4} v_{0.05}^{-3}
\left(\frac{V_0}{v}\right)\,\,{\rm s}, 
\end{align}
(note that $v_{0.05}=V_0/(0.05c)$ and not $v/(0.05c)$).
Using the diffusion timescale $t_{\rm diff}=r/v$, we can obtain the time at the peak luminosity $t_{\rm peak}$ as
\begin{align}
\begin{split}
t_{\rm peak} &\approx \sqrt{\frac{3\kappa m_{\rm ej}}{4\pi c v}}=
\sqrt{\frac{3\kappa \Sigma v}{4 c}}
\left(\frac{m_{\rm ej}}{m_{\rm a}}\right)^{1/2} t_{\rm a} \\
&= 11.5\,\Sigma_4^{1/2} v_{0.05}^{1/2} \left(\frac{v}{V_0}\right)^{1/2} \left(\frac{m_{\rm ej}}{m_{\rm a}}\right)^{1/2} \,t_{\rm a}.
\end{split}
\end{align}
This implies that the flare timescale would be comparable to the typical duration of observed QPE flares for a secondary black hole mass of $M_{\rm BH} \sim 10^4M_\odot$ if the ejected mass $m_{\rm ej}\sim m_{\rm a}$. 
However, our simulations indicate that $m_{\rm ej}$ is actually one or two orders of magnitude smaller than $m_{\rm a}$, particularly for the first (downward) outflow.
As a result, the peak time may occur by a factor of 3--10 earlier than the estimated time of $\sqrt{3 \kappa \Sigma v/4c}\,t_{\rm a}$. 

The corresponding peak luminosity is estimated to be
\begin{align}
L_{\rm peak} \approx \frac{E_{\rm ej} t_{\rm a}}{t_{\rm peak}^2} 
=\frac{2}{3}L_{\rm Edd}\left(\frac{E_{\rm ej}}{m_{\rm ej}V_0^2}\right),
\end{align}
where $L_{\rm Edd}=4\pi G c M_{\rm BH}/\kappa$ is the Eddington luminosity $\approx 1.4\times 10^{42}M_{\rm BH,4}$\,erg/s.
Since the ratio $E_{\rm ej}/(m_{\rm ej}V_0^2)$ is likely to be of order unity, the peak luminosity is always approximately equal to the Eddington luminosity regardless of the values of $V_0$ and $\Sigma$.
This suggests that the typical luminosity of observed QPEs can be reproduced for a black hole mass of $10^4$--$10^5M_\odot$.
However, for $M_{\rm BH}\agt 10^4M_\odot$, the duration of the flare by the second outflow may be longer than that of typically observed QPEs. 

If the black hole collides with a disk periodically, the outflows are driven $2\times 2$ times in each orbital period, with both downward and upward outflows present at each collision.
However, the light curve is unlikely to be simply periodic due to the distinct properties of the first (downward) and second (upward) outflows, as well as the fact that both outflows are not always observable.
In particular, the first outflow has a much lower mass than the second one, leading to a shorter flare timescale.
This means that if the disk obscures the observation of one of the two outflows, the duration of each flare would be asymmetric.
Even if both outflows can be observed (e.g., from the edge-on direction of the disk), the photon count profile of the flares should be asymmetric in the two flares observed within one orbital period. 
Nonetheless, many of the observed QPEs exhibit a photon count profile resembling two flares that do not show significant differences~\cite{2019Natur.573..381M, 2020A&A...636L...2G,2021Natur.592..704A,2024Natur.634..804N}.
This suggests that the black hole-disk collision scenario may be disfavored for interpreting these events.
A more detailed analysis of the observation scenario in the black hole-disk collision will be discussed in a separate paper. 

For a high-mass secondary black hole of mass $M_{\rm BH}$ orbiting a primary supermassive black hole of mass $M_{\rm c}$, we have to consider the gravitational radiation reaction, which has a timescale given by, e.g., \cite{1983bhwd.book.....S}
\begin{align}
\begin{split}
t_{\rm GW}&=\frac{5}{256} \frac{c^5R^4}{G^3M_{\rm c}^2 M_{\rm BH}} \\
&\approx  7.8\times 10^3\, v_{0.05}^{-8}
Q_{100} M_{\rm c,6}\,\, {\rm yr}.~~~
\end{split}
\end{align}
Here, $Q_{100}:=Q/100$, $M_{\rm c,6}:=M_{\rm c}/(10^6M_\odot)$, and we assumed a spherical orbit, i.e., $V_0=\sqrt{GM_{\rm c}/R}$, for simplicity.
As noted in \cite{2023ApJ...957...34L}, this timescale is too short to account for the coincidence with a tidal-disruption associated QPE for $M_{\rm c} \sim 10^6M_\odot$ and $R \sim 100 \,R_{\rm g}$, given that the typical event rate of tidal disruption is $10^{-5}$/yr/galaxy.
However, for $R \agt 10^3\,R_{\rm g}$, the gravitational radiation timescale $t_{\rm GW}$ becomes long enough.
In this scenario, the orbital period of a black hole orbiting a supermassive black hole is
\begin{align}
P_{\rm orb} \approx 2\pi\sqrt{\frac{R^3}{GM_{\rm c}}}
\approx 11\left(\frac{R}{10^3 \,R_{\rm g}}\right)^{3/2}
M_{\rm c,6}\,\,{\rm d}\,. \label{eq32}
\end{align}
This timescale is comparable to that of a recently discovered QPE source~\cite{Hernandez-Garcia:2025ruv}.
However, as this estimate shows, the intermediate-mass black hole-disk collision model is not suitable for interpreting the observed QPEs with a period of $\sim 10$\,h~\cite{2019Natur.573..381M, 2020A&A...636L...2G, 2021Natur.592..704A, 2024ApJ...965...12C}. To identify the QPE associated with the intermediate-mass black hole-disk collision, we have to search for a longer-period QPE.

The next question to consider is whether matter with a sufficiently high column density $\Sigma$ exists at a distance of $R \sim 10^3 \,R_{\rm g}$.
The prospects for finding such matter look promising since it is expected that tidal disruptions of ordinary stars by a supermassive black hole with a mass of $\sim 10^6M_\odot$ will occur at a distance of $R\sim 50 \,R_{\rm g}$.
Furthermore, many magnetohydrodynamics/viscous-hydrodynamics simulations (e.g., \cite{2016MNRAS.456.3929S, Shibata:2025ycs}) suggest that the majority of the resulting tidal debris, which is optically thick to the photon emission, is likely to be ejected outward.
Therefore, exploring X-ray QPE-type transients with long periods on the order of $10$\,d could be a promising strategy for searching for intermediate-mass black holes located around extragalactic centers. 

\subsection{Emission from the black-hole accretion flow and disk}
\label{sec4.2}

In this paper, we found that the mass accretion rate onto the black hole peaks at approximately $0.1\dot m_{\rm a}$ and gradually decreases over time, especially for velocities $V_0/c < 0.1$, where $\dot m_{\rm a}$ is defined in \cref{dotma}. 
For a velocity of $V_0/c \sim 0.05$ and a disk column density of $\Sigma \sim 10^4\,{\rm g/cm^2}$,
this peak mass accretion rate of $\sim 0.1\dot m_{\rm a}$ can significantly exceed the Eddington accretion rate of about $\sim 3\times 10^{-4} \dot m_{\rm a}$ (as noted in \cref{eq:Eddington_acc}), resulting in a super-Eddington accretion.

We also found that an accretion shock, characterized by a cone-shaped structure commonly observed in the Bondi-Hoyle-Lyttleton flow, forms around the black hole~(e.g., \cite{1989ApJ...336..313P, 1998ApJ...494..297F}).
In the vicinity of this shock, the maximum specific energy density is $\varepsilon \agt 10^{-2}c^2$ with the corresponding rest-mass density being comparable to that of the disk.
Assuming that the internal energy is dominated by the radiation energy, we have the relationship 
 $aT^4 \approx \rho\varepsilon$, and thus, we can express the temperature $T$ as
\begin{align}
T \approx 1.0\times 10^6 \rho_{-9}^{1/4} \varepsilon_{0.01}^{1/4}\,\,{\rm K},
\label{eq:shtemp}
\end{align}
where $\rho_{-9}:=\rho/(10^{-9}\,\mathrm{g/cm^3})$ and $\varepsilon_{0.01}:=\varepsilon/(0.01c^2)$.
The hot region surrounding the black hole is in general of the same scale as the accretion radius $r_{\rm a}$, while the high-density region within the accretion shock is confined approximately to a radius $r_{\rm sh}\alt 100\,r_{\rm g} \sim r_{\rm a}/10$.
The visible region is expected to be located outside the accretion shock because the optical thickness of the hot region is much larger than unity for the parameters considered in this paper (i.e., a maximum density of $\rho_0 \sim 10^{-8}$--$10^{-9}\,{\rm g/cm^3}$, a secondary black hole mass of $M_{\rm BH} \sim 10^4M_\odot$, and a velocity of $V_0/c \sim 0.05$).

The location of the photosphere, $r_{\rm ph}$, is determined by the detailed structure outside the accretion shock.
In the following, we parametrize the shock radius as $r_{\rm sh}=\eta r_{\rm a}$ and the photosphere radius as $r_{\rm ph}=f r_{\rm sh}$.
Note that we have $f \geq 1$ by construction and $\eta$ in the order of $\mathcal{O}(0.1)$ as suggested by our numerical results.
For a specific case where the high-density ($\rho=\rho_{\rm sh}$) region within the accretion shock is surrounded by a hotter region with a density profile decreasing as $\propto r^{-n-1}$, we can estimate the optical thickness of the hot region as
\begin{align}\label{eq34}
\begin{split}
\tau &=\displaystyle\int_{r_{\rm sh}}^{r_{\rm ph}}\kappa \rho dr \\
&\approx {\tilde \kappa} \rho_{\rm sh} r_{\rm sh} \\
&\approx  71 {\tilde \kappa}_{0.2} \eta_{0.3} M_{\rm BH,4} v_{0.05}^{-2} \rho_{{\rm sh},-9}, 
\end{split}
\end{align}
where ${\tilde \kappa}:=\kappa/n$ is the effective opacity considering the density structure with ${\tilde \kappa}_{0.2}:={\tilde \kappa}/(0.2\,{\rm cm}^2/{\rm g})$, $\rho_{{\rm sh},-9}:=\rho_{\rm sh}/(10^{-9}\,{\rm g}/{\rm cm}^3)$, and $\eta_{0.3}:=\eta/0.3$.
The parameter $f$ is then given by
\begin{equation}
f\approx \left({\tilde \kappa} \rho_{\rm sh} r_{\rm sh}\right)^{1/n}\approx \tau^{1/n}.
\end{equation}
We note that for $n \gg 1$, we have $f \approx 1$, while for small values of $n$, $f \gg 1$; e.g., for $n=1$, $f=\tau \gg 1$ as shown in \cref{eq34}.
Thus, for small values of $n$, i.e., for shallower density gradients, $f$ is larger. 

The diffusion time of photons generated in the accretion shock is given by 
\begin{align}
\begin{split}
t_{\rm diff} &\sim \tau r_{\rm ph} c^{-1} \\
&\approx 8.4\times 10^{2}  {\tilde \kappa}_{0.2} f \eta_{0.3}^2 M_{\rm BH,4}^{2} v_{0.05}^{-4} \rho_{{\rm sh},-9}\,\,{\rm s}. 
\end{split}
\end{align}
On the other hand, the accretion time scale to the secondary black hole can be estimated by $t_{\rm acc} \sim r_\mathrm{a}/V_0$, which yields
\begin{equation}
t_{\rm diff}/t_{\rm acc}\approx  1.1f \eta_{0.3}^2 {\tilde \kappa}_{0.2} M_{\rm BH,4} v_{0.05}^{-1} \rho_{{\rm sh},-9}. 
\label{eq370}
\end{equation}
For simplicity, we consider the case in which the diffusion timescale is comparable or shorter than the peak timescale of the accretion, by which we expect the radiation fields to settle into a quasi-steady profile.
Under this assumption, the luminosity can be approximated by 
\begin{align}\label{eq:luminosity-hot-region}
\begin{split}
L &\sim \frac{\pi r_{\rm ph}^2 a c T^4}{\tau} \\
&\sim 1.5\times 10^{42} {\tilde \kappa}_{0.2}^{-1} f ^2 \eta_{0.3} M_{\rm BH,4} v_{0.05}^{-2} \varepsilon_{0.01} \,{\rm erg/s}\\
&\sim 1.1 {\tilde \kappa}_{0.2}^{-1} f^2 \eta_{0.3}  v_{0.05}^{-2} \varepsilon_{0.01} L_{\rm Edd}.
\end{split}
\end{align}
This expression indicates that the luminosity $L$ could be higher for higher values of $f$, i.e., for shallower density gradients, which is consistent with the expectation that the high photon luminosity would lead to an expanded matter distribution. 

Assuming that photons are sufficiently thermalized at the photosphere, the photon temperature can be described using the effective temperature which is broadly estimated by $T_{\rm eff}\sim T/\tau^{1/4}$ as
\begin{equation}
T_{\rm eff} \sim 3.6 \times 10^5 {\tilde \kappa}_{0.2}^{-1/4} \eta_{0.3}^{-1/4} M_{\rm BH,4}^{-1/4} v_{0.05}^{1/2} \varepsilon_{0.01}^{1/4}\,{\rm K}\,.
\end{equation}
This suggests that the accretion shock is bright in the ultraviolet or soft X-ray wavelengths.

Equation~\eqref{eq:luminosity-hot-region} implies that as long as the density profile outside the accretion shock declines steeply and the photosphere is located close to the accretion shock (where $f\approx 1$), the luminosity will be comparable to the Eddington luminosity as $\varepsilon \sim V_0^2$.
On the other hand, the bolometric luminosity obtained in \cref{eq:luminosity-hot-region} is much larger than the Eddington luminosity when $f \gg 1$. For such a case, the diffusion timescale $t_{\rm diff}$ would be much longer than $t_{\rm acc}$ (cf.~Eq.~\eqref{eq370}), and thus, photons would be trapped in $r < r_{\rm ph}$, possibly forming an expanding bubble.
Notably, during the early epoch, the specific internal energy $\varepsilon$ in the vicinity of the shock can surpass $10^{-2}c^2$, leading to a luminosity that can also exceed the Eddington limit even for smaller values of $f$.
In both cases, the density profile outside the accretion shock will be highly dynamical and modified from what we obtained in our simulations due to radiative transfer effects and radiation feedback on matter.

However, $L$ cannot be unlimitedly large. 
Since the internal energy injection in the accretion shock can be broadly estimated by 
\begin{equation}
\sim 0.1 {\dot m}_{\rm a}V_0^2\approx 1.3\times 10^{43}\,M_{\rm BH,4} v_{0.05} \Sigma_{4} \,{\rm erg/s}, 
\end{equation}
the bolometric luminosity may be limited by this value. The radiative cooling will also play an important role in determining the structure inside the hot region for a large value of $f$.

We note that, while it is often assumed that actual luminosity is restricted by the Eddington luminosity, this limit can be exceeded as the matter profile is non-spherically symmetric and highly dynamical, as illustrated in, e.g., \cref{fig1}.
Indeed, numerical simulations of super-Eddington accretions in black hole-torus systems (e.g.,~\citep{Ohsuga:2005bh,Ohsuga:2007ba}) illustrate that the bolometric luminosity can exceed the Eddington limit by a factor of up to $\sim 10$, depending on the viewing angle. 
If this occurs, the peak luminosity of the accretion-powered emission could be sufficiently high to explain observations of the bright class of QPEs associated with a secondary black-hole mass of $\sim 10^{4}\,M_\odot$ (as discussed below).
However, a quantitative assessment of this possibility necessitates simulations that incorporate both radiative transfer and radiation feedback to matter.

Although the accretion disk is not formed around the secondary black hole in the current setting where axisymmetry is assumed, it is reasonable to expect that such a formation could occur in reality, e.g., for model \texttt{M050.1000.10}.
This expectation arises because the disk surrounding the supermassive black hole should exhibit Keplerian velocity, and the typical specific angular momentum $j$ of the disk matter in the frame comoving with the secondary black hole would be 
\begin{equation}
j = \frac{r_{\rm a}}{2R}V_0 \times r_{\rm a}=0.4 Q_{100}^{-1}
v_{0.05}^{-1}
\frac{GM_{\rm BH}}{c}, 
\end{equation}
which can exceed $GM_{\rm BH}/c$ if the inverse mass ratio $Q^{-1}=M_{\rm BH}/M_{\rm c}$ is not extremely small or the velocity $V_0/c$ is small.
If this condition is satisfied, an accretion disk is likely to form around the secondary black hole, resulting in emissions approximately at the Eddington luminosity during the super-Eddington mass accretion phase, where $\dot m_{\rm BH} > \dot m_{\rm Edd}$ (see, for instance, \cite{2005ApJ...628..368O, 2014ApJ...796..106J, 2014MNRAS.439..503S, 2016MNRAS.459.3738I, 2016MNRAS.456.3929S, 2018PASJ...70..108K, 2021PASJ...73..450K, 2024MNRAS.532.4826T}).
To investigate this possibility, we plan to conduct a non-axisymmetric simulation and/or an axisymmetric simulation that incorporates the angular momentum of the disk.

Regardless of the formation of the accretion disk, the duration of the bright emission is governed by the accretion timescale, which is as long as $10^6 \,t_{\rm g}\approx 5 \times 10^4M_{\rm BH,4}\,{\rm s}$ for $V_0/c \alt 0.1$ as indicated by our simulations.
This timescale is significantly longer than $t_{\rm a}$ and $t_{\rm peak}$ discussed in \cref{sec4.1}.
Since the peak luminosity comparable to or exceeding the Eddington limit $\agt L_{\rm Edd}$ can be maintained for a considerably long duration of $10^6t_{\rm g}$, which is comparable to the orbital period particularly for small mass ratio $Q$, this long-term mass accretion is likely the primary emission source for black hole-collision events.
As a result, the light curves observed in the black hole-disk collision events may differ significantly from those of the sharp peaks seen in QPEs, especially for a secondary black hole with mass $M_{\rm BH} \agt 10^4M_\odot$.
Instead of sharp peaks, the light curves might display a sawtooth pattern.
Therefore, to effectively discover intermediate-mass black holes in galactic centers through transient searches, it is advisable to target events with recurrence times that are comparable to or constitute a significant fraction of the time duration of the eruption.
Conversely, for mass ratio $Q\sim 100$, the ratio of the duration to the recurrence time is similar to that of the typically observed QPEs~\cite{2024Natur.634..804N,  2023A&A...675A.152Q, Chakraborty:2025ntn, Hernandez-Garcia:2025ruv}.

A recent paper~\cite{Hernandez-Garcia:2025ruv} reports on a QPE event characterized by a long recurrence time of $\sim 4.5$\,d and a duration of $\sim 1.5$\,d for each flare.
In a scenario involving disk-star (or black hole) collision, the estimated orbital period of the star (or secondary black hole) would be $\sim 9$\,d.
For this system, the mass of the primary black hole $M_c$ is estimated to be $\sim 10^6M_\odot$.
Consequently, the orbital velocity can be estimated using \cref{eq32} to be $\sim 0.034c$ assuming a circular orbit.
If we further assume the mass of the secondary black hole to be $M_{\rm BH}\sim 2 \times 10^4M_\odot$ with this orbital velocity, the accretion timescale onto the secondary black hole with the super-Eddington accretion would be $\sim 1$--$2\times 10^6t_{\rm g}$, which translates to about 1--2\,d.
For this event, the peak luminosity is estimated to be $\sim 2 \times 10^{43}$\,erg/s~\cite{Hernandez-Garcia:2025ruv}.
Therefore, if the luminosity during the super-Eddington accretion phase exceeds several times the Eddington luminosity $L_{\rm Edd}$, this event might be interpreted as the result of repeated collisions between an intermediate-mass black hole of mass $\sim 2 \times 10^4M_\odot$ and a transiently formed disk.
To scrutinize the validity of this hypothesis, we require a detailed radiation hydrodynamics study to predict the light curve and spectrum during the process of dynamical matter accretion onto the black hole. 

In this scenario, there is a binary black hole, which is expected to emit strong gravitational waves.
However, the frequency of $\sim 2/(9\,{\rm d}) \approx 2.6\times 10^{-6}$\,Hz unfortunately falls outside the frequency bands detectable by both LISA and pulsar-timing arrays.
Despite this limitation, detecting such a binary system through electromagnetic signals will be useful for estimating the event rate of this type of binary for LISA.

\subsection{Effect of the emission from the accretion disk of the primary black hole}

The transient emissions analyzed in this section may occur on top of a background emission from the accretion disk surrounding the primary massive black hole.
In the absence of interactions with a secondary black hole, this disk could remain in a quasi-steady state, with its predominant emissions originating near the innermost stable circular orbit. If the disk is formed as a result of a tidal disruption of an ordinary star, the mass accretion onto the primary black hole is initially in the super-Eddington regime (e.g., \cite{Rossi} for a review), and thus, the luminosity would be approximately equal to the Eddington luminosity, which should be by a factor of $\sim M_{\rm c}/M_{\rm BH}$ higher than the luminosity associated with the collision events.
The standard accretion disk scenario also indicates that the typical temperature of the accretion disk of the primary black hole would be in the soft X-ray region ($10^6$--$10^7$\,K) for masses $M_{\rm c} \sim 10^6M_\odot$~\cite{shakura1973black, Novikov:1973, 1983bhwd.book.....S}.
Therefore, both the flares and the emissions from the accretion flow onto the secondary black hole could be significantly dimmer than the disk emissions (for $Q\gg 1$).
Thus, the emissions associated with the collision events may only be observable when the emission from the disk of the primary massive black hole becomes less bright, i.e., after the mass accretion rate onto the primary black hole becomes sufficiently small, although a high column density near the collision region is still necessary.
Alternatively, if the emission from the primary black hole is obscured by the surrounding matter, which may form after the viscous evolution of the disk~\cite{Shibata:2025ycs}, the emission from the secondary black hole could be detectable~\cite{2023ApJ...957...34L}.
In both scenarios, a special timing appears to be essential for observing the emission from the collision (see also~\cite{linial2024ultraviolet} for the observability of the flares in the context of star-disk collisions).

\section{Summary}\label{sec6}

We conducted a general relativistic simulation of a collision between a black hole and a disk using our FMR code, {\tt SACRA-2D}, assuming that both the black hole and the disk are orbiting a supermassive black hole with orthogonal orbits.
The simulation was carried out for a variety of relative velocities between the black hole and the disk, as well as the thickness of the disk.
Our disk model was designed to achieve a realistic force balance between the supermassive black hole and the gas pressure when located far from the black hole.
Our findings are as follows:
\begin{itemize}
\item When the black hole collides with and passes through the disk, outflows are driven as found in \cite{1998ApJ...507..131I}.
However, the mass of the outflows, particularly for the first outflow, is smaller than $m_{\rm a}$, contrary to previous expectations, while the energy associated with the outflows can be expressed as $\sim m_{\rm ej}V_0^2$.
Consequently, the flares associated with this mass ejection would shine with the Eddington luminosity of the black hole, occurring on a timescale shorter than previously anticipated.
\item After the black hole passes through the disk, mass inflow onto the black hole continues for a long timescale, especially for small values of velocity $V_0/c \alt 0.1$.
For plausible values of the column density of the disk, the mass accretion rate exceeds the Eddington limit.
When the velocity $V_0/c \alt 0.1$, this super-Eddington accretion, which is likely to produce Eddington-luminosity emission, would continue for a timescale $\sim 10^6 \,t_{\rm g}$.
This emission timescale could be comparable to the orbital period of the black hole around the more massive primary black hole for a small mass ratio of order $Q=\mathcal{O}(10)$.
In such cases, the emission may not be recognized as a QPE type.
On the other hand, for mass ratio $Q \sim 100$, the typical recurrence time relative to the duration of the observed QPEs is likely to match, reproducing the observed results. 
\item An accretion shock with a high temperature is formed due to the mass infall toward the black hole.
For velocity $V_0/c \alt 0.1$, convection takes place upstream of the shock wave.
With plausible values of the column density of the disk, the shock region and convective area are likely to shine with nearly Eddington luminosity.
The emitted wavelength would be in the ultraviolet or soft X-ray bands. 
\item Assuming a supermassive black hole of mass $M_{\rm c}\sim 10^6M_\odot$ and a secondary black hole of mass $M_{\rm BH}\sim 2 \times 10^4M_\odot$ with an orbital radius of $R \sim 10^3R_{\rm g}$, the latest event of a QPE~\cite{Hernandez-Garcia:2025ruv} may be explained by the intermediate-mass black hole-disk collision hypothesis. 
\end{itemize}

In this paper, we do not provide detailed analyses of the electromagnetic emissions.
A more thorough examination is necessary to predict the expected wavelengths of the electromagnetic signals, which we plan to address in a future paper.
Another issue we intend to explore in the future is conducting three-dimensional simulations.
This is important because (i) the accretion disk in reality rotates around a primary supermassive black hole, resulting in orbital velocity as well as density and velocity gradients along the disk plane, and (ii) the orbit of the secondary black hole is generally not perpendicular to the disk plane.
We aim to perform these three-dimensional simulations in our subsequent work.

\acknowledgements

We thank Re'em Sari for an educational explanation of QPE to us. We also thank Kenta Hotokezaka and Tatsuya Matsumoto for useful discussions. Numerical computation was performed on the cluster Sakura at the Max Planck Computing and Data Facility. This work was in part supported by Grant-in-Aid for Scientific Research (grant No.~23H04900) of Japanese MEXT/JSPS.

\appendix

\begin{figure*}
    \centering
    \includegraphics[width=\textwidth]{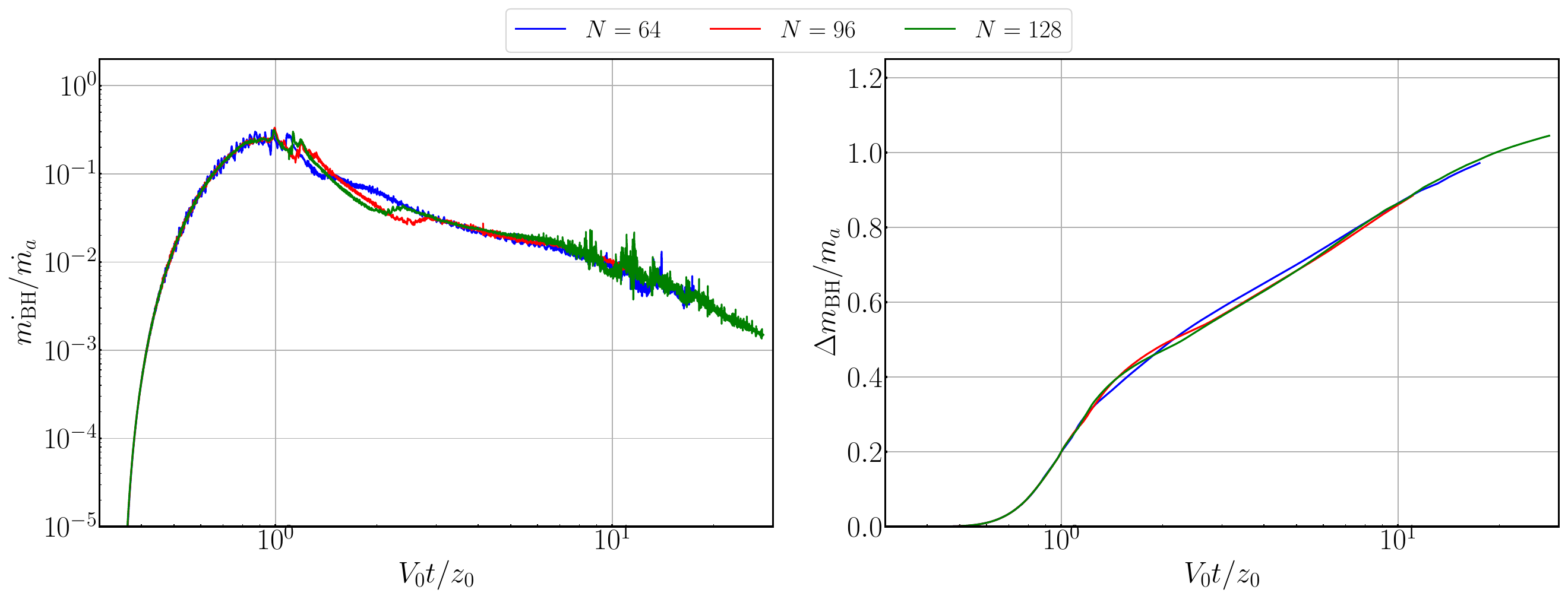}
    \caption{Evolution of $\dot m_{\rm BH}$ and $\Delta m_{\rm BH}$ for model \texttt{M050.1000.10} with three different grid resolutions, $N=64, 96$, and $128$.}
    \label{flux_conv}
\end{figure*}

\begin{figure*}
    \centering
    \includegraphics[width=\textwidth]{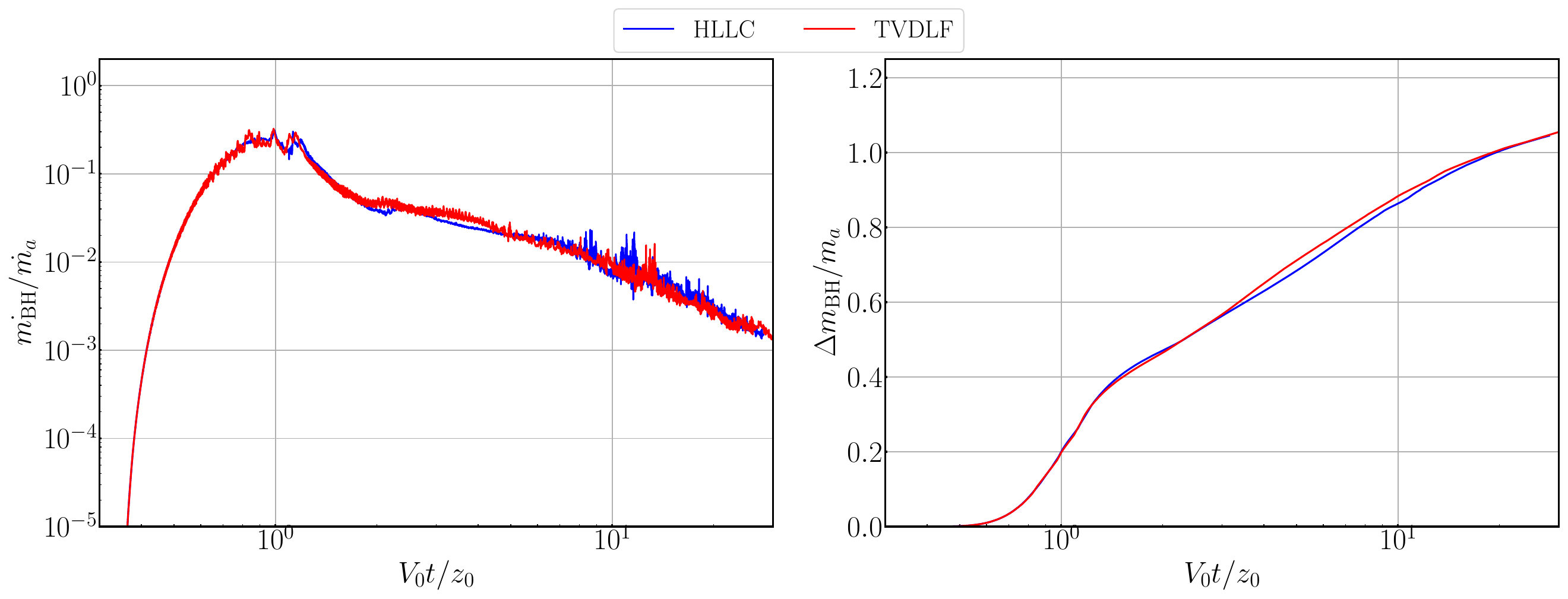}
    \caption{Evolutions of $\dot m_{\rm BH}/\dot m_a$ and $\Delta m_{\rm BH}/m_a$ for model \texttt{M050.1000.10} with HLLC and TVDLF solvers.}
    \label{flux_tvdlf}
\end{figure*}

\section{Convergence test} \label{app1}
To examine how the numerical results depend on grid resolution, we simulate model \texttt{M050.1000.10} using the HLLC solver with three different grid resolutions $N=128$, $96$, and $64$.
The other FMR settings remain the same across all three simulations, except for the grid resolution. 
Figure~\ref{flux_conv} presents the results of $\dot m_{\rm BH}$ and $\Delta m_{\rm BH}$ for the three grid resolutions with the HLLC solver.
This figure demonstrates that the evolution process is only weakly dependent on the grid resolution for $N \geq 64$, although the short-term evolution is influenced by the resolution.
This behavior is expected, as shocks, which reduce convergence to first order in shock-capturing hydrodynamics schemes, are formed frequently around the black hole in this problem.
However, this does not significantly affect the overall evolution of the system. 
One noteworthy point is that with improved grid resolution, the oscillation of $\dot m_{\rm BH}$ during the early times with $t \alt z_0/V_0$ is suppressed, and the late-time convection appears earlier.

\section{Comparison of Riemann solvers} \label{app2}
To investigate how the numerical results depend on the chosen Riemann solver, we also simulate model \texttt{M050.1000.10} using the TVDLF solver under the same resolution $N=128$ and grid settings.
Figure~\ref{flux_tvdlf} compares the mass accretion rate $\dot m_{\rm BH}$ and total accreted mass $\Delta m_{\rm BH}$ derived from the two solvers.
For the TVDLF solver, the oscillation of the mass accretion rate $\dot m_{\rm BH}$ in the early stages with $t \alt z_0/V_0$ is slightly enhanced.
However, the influence of the late-time convection is reduced, resembling a lower resolution effectively.

\begin{figure*}
    \centering
    \includegraphics[width=\textwidth]{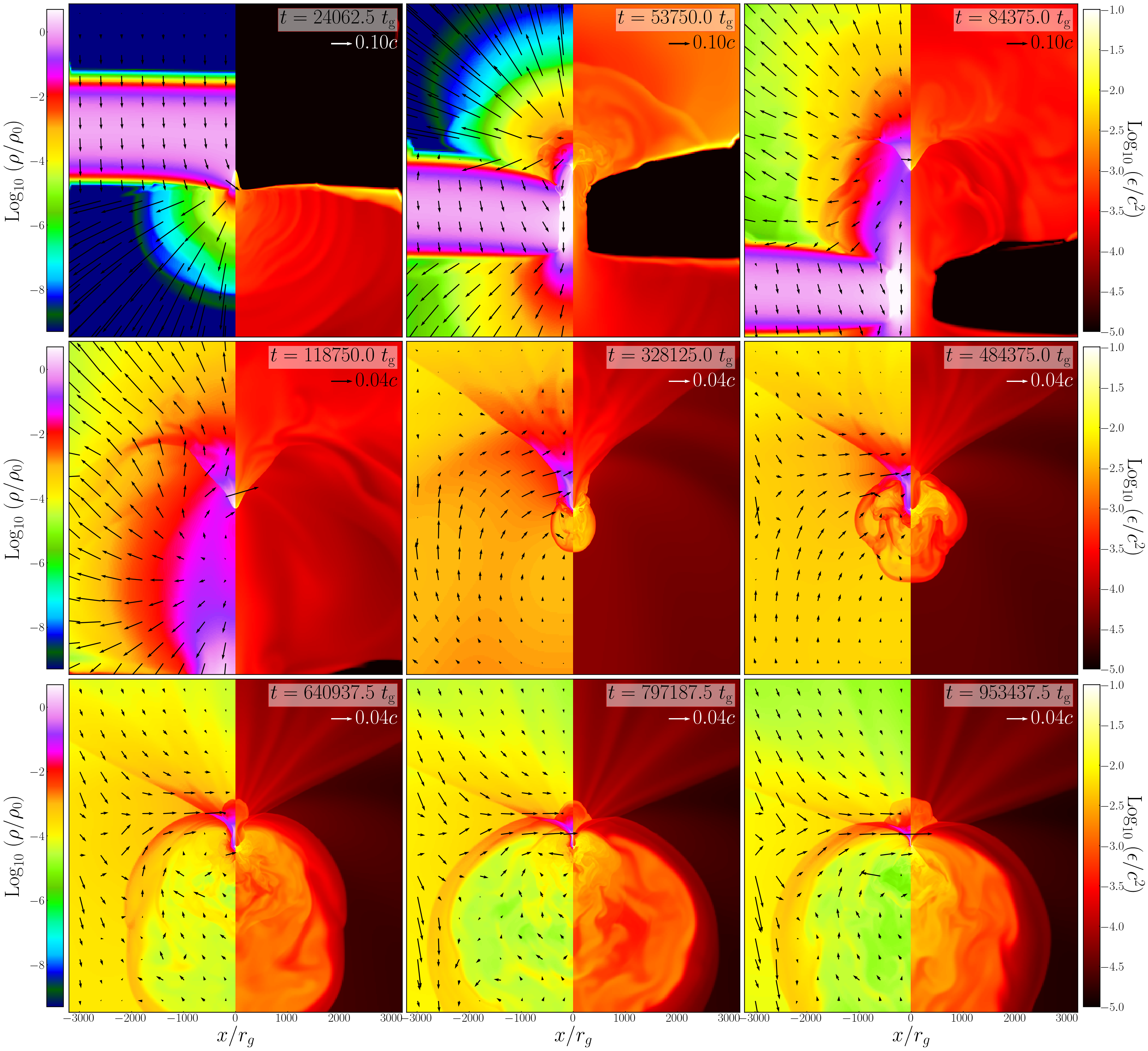}
    \caption{The same as \cref{fig1} but with the TVDLF solver.}
    \label{fig1_TVDLF}
\end{figure*}

Figure~\ref{fig1_TVDLF} is the same as \cref{fig1}, i.e., the plots of the rest-mass density and specific internal energy at identical time slices.
We observed that for the first five snapshots, these plots are nearly indistinguishable from those in \cref{fig1}, indicating that the numerical results show weak dependence on the selected Riemann solvers.
In the later stages, the convective motion appears to be slightly suppressed with the TVDLF solver, but no striking difference is seen. 
Overall, both Riemann solvers yield essentially the same results.

\bibliographystyle{apsrev4-2}
\bibliography{reference}

\end{document}